\documentclass{article}

\usepackage{amsmath,amssymb}
\usepackage{mathrsfs}

\usepackage{hyperref}
\usepackage{bbm}
\usepackage[paper=a4paper,text={17cm,24cm},centering]{geometry}
\usepackage{graphicx}
\usepackage{subcaption}
\usepackage{adjustbox}
\usepackage{enumerate}
\usepackage{dsfont}
\usepackage{physics}
\usepackage{mdframed}

\usepackage{tikz}
\usetikzlibrary{arrows.meta,arrows}
\usepackage[compat=1.1.0]{tikz-feynman}

\usepackage{authblk}
\usepackage{titlesec}
\titleformat{\paragraph}[block]{\normalfont\normalsize\bfseries}{\theparagraph}{1em}{}

\def\be{\begin{equation}}
\def\ee{\end{equation}}
\def\MSbar{\overline{\text{MS}}}
\DeclareMathOperator{\sgn}{sgn}
\DeclareMathOperator{\atanh}{atanh}

\title{Extracting light-cone wave functions from covariant amplitudes:\\ a detailed study in scalar field theory}
\author{Stéphane Munier}
\affil{\it\normalsize CPHT, CNRS, École polytechnique, Institut Polytechnique de Paris, 91120 Palaiseau, France}
\date{December 26, 2025; Latest version: July 14, 2026}

\begin{document}

\maketitle

\begin{abstract}
We propose a conjectured formula that systematically maps covariant off-shell amplitudes to light-cone wave functions in scalar field theory. Through an explicit comparison at one-loop accuracy, we establish its equivalence to the light-cone perturbation theory series, thereby validating the conjecture at this order. Applying this formula, we efficiently re-derive wave functions from known covariant amplitudes, bypassing both the conceptual complexities of light-cone quantization and the technical challenges of perturbative calculations in this framework. In addition to simplifying computations, this approach opens new avenues for applications in gauge theories and deeper explorations of the fundamental equivalence between covariant and light-cone quantization.
\end{abstract}

\def\onetooneoneloop{\scalebox{1.}{\begin{tikzpicture}[baseline={(current bounding box.center)}]
  \begin{feynman}
  \draw[-] (-0.5,0) to (0.5,0);
  \node at (0,0.25) {${k}$};
  \draw(1,0) circle [radius=0.5];
  \draw[-](1.5,0) to (2.5,0);
  \node at (1,0.75) {${l}$};
  \node at (1,-0.7) {${k}-{l}$};
  \end{feynman}
\end{tikzpicture}}}

\def\onetooneoneloopct#1#2#3{\scalebox{1.}{\begin{tikzpicture}[baseline={(current bounding box.center)}]
  \begin{feynman}
  \draw[-](-0.1,0) to (0.85,0);
  \node at (0.4,0.25) {#1};
  \draw(1,0) circle [radius=0.15];
  \draw(0.9,-0.1) to (1.1,0.1);
  \draw(0.9,0.1) to (1.1,-0.1);
  \node at (1, 0.3) {\tiny #2};
  \node at (1,-0.3) {\phantom{\tiny #2}};
  \draw[-](1.15,0) to (2,0);
  \node at (1.5,0.3) {#3};
  \end{feynman}
\end{tikzpicture}}}

\def\triangle{\scalebox{1.}{\begin{tikzpicture}[baseline={(current bounding box.center)}]
  \begin{feynman}
  \draw[-] (-0.5,0) to (0.5,0);
  \node at (0,0.25) {${k}$};
  \draw[-](0.5,0) to (1.5,.8);
  \draw[-](0.5,0) to (1.5,-.8);
  \draw[-](1.5,0.8) to (1.5,-.8);
  \draw[-](1.5,0.8) to (2.5,.8);
  \draw[-](1.5,-0.8) to (2.5,-.8);
  \node at (.6,0.7) {\small $k_1-l$};
  \node at (.6,-0.7) {\small $k_2+l$};
  \node at (1.7,0) {$l$};
  \node at (2.,1.05) {${k_1}$};
  \node at (2.,-1.1) {$k_2$};
  \end{feynman}
\end{tikzpicture}}}

\def\trianglect{\scalebox{1.}{\begin{tikzpicture}[baseline={(current bounding box.center)}]
  \begin{feynman}
    \node at (-0.05,-0.2) {${k}$};
  \draw[-](-0.5,-0.5) to (0.35,-0.5);
  \draw(0.5,-0.5) circle [radius=0.15];
  \draw(0.6,-0.6) to (0.4,-0.4);
  \draw(0.6,-0.4) to (0.4,-0.6);
  \draw(1.15,0.3) arc (90:180:0.65);
  \draw(1.15,-1.3) arc (270:180:0.65);
  \draw[-] (1.15,0.3) to (2,0.3);
  \draw[-] (1.15,-1.3) to (2,-1.3);
 \node at (1.5,.55) {${k_1}$};
  \node at (1.5,-1.6) {$k_2$};
  \end{feynman}
\end{tikzpicture}}}

\newcommand{\contourVertex}{
  \begin{tikzpicture}[>=Stealth, scale=1.0]
    \draw[->, thick] (-5.5,0) -- (4.5,0) node[right] {$\Re(\delta_1^-)=\Re(k_1^-)-\tilde k_1^-$};
    \draw[->, thick] (0,-1.2) -- (0,1.2) node[above] {$\Im(\delta_1^-)=\Im(k_1^-)$};

    \coordinate (pole1) at (-1.2,0.8);
    \draw[line width=0.8pt] ($(pole1)+(-0.12,-0.12)$) -- ($(pole1)+(0.12,0.12)$);
    \draw[line width=0.8pt] ($(pole1)+(-0.12,0.12)$) -- ($(pole1)+(0.12,-0.12)$);
    \draw[dashed] (pole1) -- (-1.2,0);
    \node[below] at (-1,0) {\footnotesize$\Delta_R(\vec k_1,\vec k_2)$};
    
    \coordinate (cut1) at (-2.5,0.8);
    \draw[thick, decorate, decoration={snake, amplitude=0.6mm, segment length=4mm}]
      (cut1) -- (-5.5,0.8) node[left] {\footnotesize $-\infty+i0^+$};    

    \draw[dashed] (cut1) -- (-2.5,0);
    \node[below] at (-3.8,0) {\footnotesize$\Delta_R(\vec k_1,\vec k_2)-3m_R^2/(2k_2^+)$};
      
    \coordinate (pole2) at (0,-0.8);
    \draw[line width=0.8pt] ($(pole2)+(-0.12,-0.12)$) -- ($(pole2)+(0.12,0.12)$);
    \draw[line width=0.8pt] ($(pole2)+(-0.12,0.12)$) -- ($(pole2)+(0.12,-0.12)$);

    \coordinate (cut2) at (1.1,-0.8);
    \draw[dashed] (cut2) -- (1.1,0);
    \node[above] at (1.3,0) {\footnotesize$3m_R^2/(2k_1^+)$};

    \draw[thick, decorate, decoration={snake, amplitude=0.6mm, segment length=4mm}]
      (cut2) -- (4.5,-0.8) node[right] {\footnotesize $+\infty-i0^+$};

    \node[above left] at (0,0) {$0$};

    \begin{scope}
      \draw[dotted, line width=0.8pt, domain=0:350, variable=\t,
        postaction={decorate, decoration={markings, mark=at position 0.1 with {\arrow{<}}}}]
        plot ({0+0.3*cos(\t)}, {-0.8+0.3*sin(\t)});

      \draw[dotted, thick, <-] (4.5,-0.5) -- (1.2,-0.5);

      \draw[dotted, thick, domain=90:270, variable=\t]
        plot ({1.2+0.3*cos(\t)}, {-0.8+0.3*sin(\t)});

      \draw[dotted, thick, <-] (1.2,-1.1) -- (4.5,-1.1);
    \end{scope}
  \end{tikzpicture}
}

\newcommand{\contourSelf}{
  \begin{tikzpicture}[>=Stealth, scale=1.0]
    \draw[->, thick] (-4.5,0) -- (4.5,0) node[right] {$\Re(\delta_1^-)=\Re(k_1^-)-\tilde k_1^-$};
    \draw[->, thick] (0,-1.2) -- (0,1.2) node[above] {$\Im(\delta_1^-)=\Im(k_1^-)$};

    \coordinate (pole1) at (-1.2,0.8);
    \draw[line width=0.8pt] ($(pole1)+(-0.12,-0.12)$) -- ($(pole1)+(0.12,0.12)$);
    \draw[line width=0.8pt] ($(pole1)+(-0.12,0.12)$) -- ($(pole1)+(0.12,-0.12)$);
    \draw[dashed] (pole1) -- (-1.2,0);
    \node[below] at (-1.5,0) {\footnotesize$\Delta_R(\vec k_1,\vec k_2)$};

    \coordinate (pole2) at (0,-0.8);
    \draw[line width=0.8pt] ($(pole2)+(-0.12,-0.12)$) -- ($(pole2)+(0.12,0.12)$);
    \draw[line width=0.8pt] ($(pole2)+(-0.12,0.12)$) -- ($(pole2)+(0.12,-0.12)$);
    \draw[line width=0.7pt] (pole2) circle [radius=0.18];

    \coordinate (cut2) at (1.1,-0.8);
    \draw[dashed] (cut2) -- (1.1,0);
    \node[above] at (1.3,0) {\footnotesize$3m_R^2/(2k_1^+)$};

    \draw[thick, decorate, decoration={snake, amplitude=0.6mm, segment length=4mm}]
      (cut2) -- (4.5,-0.8) node[right] {\footnotesize $+\infty-i0^+$};

    \node[above left] at (0,0) {$0$};

    \draw[dashed, line width=0.8pt, domain=0:350, variable=\t,
      postaction={decorate, decoration={markings, mark=at position 0.25 with {\arrow{>}}}}]
      plot ({-1.2+0.3*cos(\t)}, {0.8+0.3*sin(\t)});

    \begin{scope}
      \draw[dotted, line width=0.8pt, domain=0:350, variable=\t,
        postaction={decorate, decoration={markings, mark=at position 0.1 with {\arrow{<}}}}]
        plot ({0+0.4*cos(\t)}, {-0.8+0.4*sin(\t)});

      \draw[dotted, thick, <-] (4.5,-0.5) -- (1.2,-0.5);

      \draw[dotted, thick, domain=90:270, variable=\t]
        plot ({1.2+0.3*cos(\t)}, {-0.8+0.3*sin(\t)});

      \draw[dotted, thick, <-] (1.2,-1.1) -- (4.5,-1.1);
    \end{scope}
  \end{tikzpicture}
}


\def\scalarselfenergyone{%
  \begin{tikzpicture}[
    baseline={([yshift=-2.5pt]current bounding box.center)},
    scale=1.2, 
    every node/.style={transform shape=false} 
  ]
    \begin{feynman}
      \node at (-0.75,-0.2) {${k}$};
      \draw(-1.,-0.5) to (-0.5,-0.5);
      \draw(0.,0) arc (90:270:0.5);
      \node at (-0.3,0.2) {\footnotesize ${k'_1}$};
      \draw(0.3,0.) circle [radius=0.3];
      \node at (0.3,0.52) {\footnotesize ${l_1}$};
      \node at (0.3,-0.53) {\footnotesize ${l_2}$};
      \draw(0.6,0) to (1.,0);
      \draw(0.,-1) to (1.,-1);
      \node at (1.3,0) {${k}_1$};
      \node at (1.3,-1) {${k}_2$};
      \node at (-0.35,-0.5) {\footnotesize $0$};
      \node at (0.18,0.07) {\footnotesize $1'$};
      \node at (.7,0.15) {\footnotesize $1$};
    \end{feynman}
  \end{tikzpicture}
}

\def\scalarselfenergyonect{%
  \begin{tikzpicture}[
    baseline={([yshift=-2.5pt]current bounding box.center)},
    scale=1.2, 
    every node/.style={transform shape=false} 
  ]
    \begin{feynman}
      \node at (-0.25,-0.2) {${k}$};
      \node at (0.2,0.2) {\footnotesize ${k'_1}$};
      \draw(-0.5,-0.5) to (0.,-0.5);
      \draw(0.5,0) arc (90:270:0.5);
      \draw(0.65,0.) circle [radius=0.15];
      \draw(0.65-0.1,0.-0.1) to (0.65+0.1,0.+0.1);
      \draw(0.65-0.1,0.+0.1) to (0.65+0.1,0.-0.1);
      \node at (0.3,0.6) {\phantom{${k'_1}$}};
      \draw(0.8,0) to (1.3,0);
      \draw(0.5,-1) to (1.3,-1);
      \node at (1.6,0) {${k}_1$};
      \node at (1.6,-1) {${k}_2$};
    \end{feynman}
  \end{tikzpicture}
}

\def\scalarvertex{%
  \begin{tikzpicture}[
    baseline={([yshift=-2.5pt]current bounding box.center)},
    scale=1.2, 
    every node/.style={transform shape=false} 
  ]
    \begin{feynman}
      \node at (0.1,-0.2) {${k}$};
      \node at (1,0.3) {\footnotesize ${k_1'}$};
      \node at (0.5,-1.1) {\footnotesize ${k_2'}$};
      \node at (1.1,-.3) {\footnotesize ${l}$};
      \node at (1,-1.3) {\footnotesize \phantom{${k_2'}$}};
      \draw[-](0,-0.5) to (0.5,-0.5);
      \node at (0.65,-0.5) {\footnotesize $0$};
      \draw(1,0) arc (90:270:0.5);
      \draw[-] (1.5,0) to (1,-1);
      \draw[-] (1,0) to (1.8,0.0);
      \draw[-] (1,-1) to (1.8,-1);
      \node at (1.5,0.15) {\footnotesize $1$};
      \node at (1.,-1.2) {\footnotesize $2$};
      \node at (2.2,0) {$k_1$};
      \node at (2.2,-1) {$k_2$};
    \end{feynman}
  \end{tikzpicture}
}

\def\scalarvertexct{%
  \begin{tikzpicture}[
    baseline={([yshift=-2.5pt]current bounding box.center)},
    scale=1.2, 
    every node/.style={transform shape=false} 
  ]
    \begin{feynman}
      \node at (1,0.3) {\footnotesize \phantom{${k_1'}$}};
      \node at (1,-1.3) {\footnotesize \phantom{${k_2'}$}};
      \node at (-0.05,-0.2) {${k}$};
      \draw[-](-0.15,-0.5) to (0.35,-0.5);
      \draw(0.5,-0.5) circle [radius=0.15];
      \draw(0.6,-0.6) to (0.4,-0.4);
      \draw(0.6,-0.4) to (0.4,-0.6);
      \draw(0.85,0) arc (90:180:0.35);
      \draw(0.85,-1) arc (270:180:0.35);
      \draw[-] (0.85,0) to (1.8,0.0);
      \draw[-] (0.85,-1) to (1.8,-1);
      \node at (2.2,0) {$k_1$};
      \node at (2.2,-1) {$k_2$};
    \end{feynman}
  \end{tikzpicture}
}

\def\scalarselfenergytwo{\scalebox{1.}{\begin{tikzpicture}[baseline={([yshift=-2.5]current bounding box.center)}]
  \begin{feynman}
    \draw(-1.,-1) node[left] {$ k$} to (0,-1);
    \draw(1.,0.) arc (90:270:1);
    \node at (0.,0.0) {${k'_1}$};
    \node at (0.,-2.0) {${k'_2}$};
    \draw(1.5,0.) circle [radius=0.5]; 
    \node at (1.8,0.4) [above]{$ l_1$};
    \node at (1.2,-0.4) [below]{$ l_1'$};
    \draw(2,0) to (3.,0) node[right]{$ k_1$};
    \draw(2.,-2) to (3.,-2) node[right]{$ k_2$};
    \node at (1.8,-1.6) [above] {${l}_2$};
    \node at (1.2,-2.4) [below] {$ l_2'$};
    \draw(1.5,-2) circle [radius=0.5];
    \node at (0,-1)[right] {\footnotesize $0$};
    \node at (1,0)[right] {\footnotesize $1'$};
    \node at (2,0)[left] {\footnotesize $1$};
    \node at (1,-2)[right] {\footnotesize $2'$};
    \node at (2,-2)[left] {\footnotesize $2$};
    \end{feynman}
\end{tikzpicture}}}

\def\LCtriangle{%
  \begin{tikzpicture}[
    baseline={([yshift=-2.5pt]current bounding box.center)},
    scale=1.5, 
    every node/.style={transform shape=false} 
  ]
    \begin{feynman}
      \node at (-0.1,-0.2) {\footnotesize $\vec k_1+\vec k_2$};
      \node at (0.5,0.2) {\footnotesize $\vec k_1+\vec l$};
      \node at (1,-1.2) {\footnotesize $\vec k_2-\vec l$};
      \node at (1.4,-0.3) {\footnotesize $\vec{l}$};
      \draw[-] (0,-0.5) to (0.5,-0.5);
      \draw (1,0) arc (90:270:0.5);
      \draw[-] (1,0) to (1.5,-1);
      \draw[-] (1,0) to (1.8,0.0);
      \draw[-] (1,-1) to (1.8,-1);
      \node at (2.2,0) {$\vec k_1$};
      \node at (2.2,-1) {$\vec k_2$};
    \end{feynman}
  \end{tikzpicture}
}

\def\LCtrianglebis{%
  \begin{tikzpicture}[
    baseline={([yshift=-2.5pt]current bounding box.center)},
    scale=1.5, 
    every node/.style={transform shape=false} 
  ]
    \begin{feynman}
      \node at (-0.1,-0.2) {\footnotesize $\vec{k}_1+\vec k_2$};
      \node at (1,0.2) {\footnotesize $\vec k_1 -\vec l$};
      \node at (0.5,-1.2) {\footnotesize $\vec k_2+\vec l$};
      \node at (1.1,-0.3) {\footnotesize $\vec{l}$};
      \draw[-] (0,-0.5) to (0.5,-0.5);
      \draw (1,0) arc (90:270:0.5);
      \draw[-] (1.5,0) to (1,-1);
      \draw[-] (1,0) to (1.8,0.0);
      \draw[-] (1,-1) to (1.8,-1);
      \node at (2.2,0) {$\vec k_1$};
      \node at (2.2,-1) {$\vec k_2$};
    \end{feynman}
  \end{tikzpicture}
}

\newcommand{\uvplane}{
\begin{tikzpicture}[scale=3]

\def\wzero{0.45}
\def\muzero{0.55}

\draw (0,1) -- (\wzero,\muzero);
\draw (0,0) -- (\wzero*1.5,\muzero*1.5) node[above right]{$k_1^2 u+k_2^2 v=0$};
\draw (\wzero,\muzero) -- (1,0);

\fill[gray!20]
  (0,0) -- (0,1) -- (\wzero,\muzero) -- cycle;
\fill[pattern=dots]
  (0,0) -- (1,0) -- (\wzero,\muzero) -- cycle;

\draw[dashed] (0,\muzero) -- (\wzero,\muzero);
\draw[dashed] (\wzero,0) -- (\wzero,\muzero);

\node[left] at (0,1) {$1$};
\node[left] at (0,\muzero) {$v_0$};
\node[below] at (\wzero,0) {$u_0$};
\node[below] at (1,0) {$1$};

\fill (\wzero,\muzero) circle (0.5pt);

\draw[->] (0,0) -- (1.2,0) node[right] {$u$};
\draw[->] (0,0) -- (0,1.2) node[above] {$v$};

\end{tikzpicture}
}


\section{Introduction}
\label{sec:introduction}

In perturbative quantum field theory, the covariant formulation of scattering amplitudes is the standard approach for making quantitative predictions in phenomenological applications. However, processes such as the high-energy scattering of an electron or proton off a large nucleus resemble the nearly instantaneous scattering of a quantum particle from a classical external field. In such cases, \(S\)-matrix elements can be expressed as linear combinations of matrix elements of tensor products of Wilson lines, representing the eikonal scattering of free, independent partons at very high energies off the external field. The weights of these partonic configurations, determined by the distribution amplitudes of the scattering partons in the physical initial and final states, are encoded in light-cone wave functions (see Ref.~\cite{Kovchegov:2012mbw} for a textbook and Ref.~\cite{Angelopoulou:2023qdm} for a recent review).

Computing these wave functions requires quantizing the theory in a non-Lorentzian light-cone frame~\cite{Weinberg:1966jm,Kogut:1969xa,Bjorken:1970ah,Lepage:1980fj,Pauli:1985pv, Brodsky:1997de}, a procedure that introduces both conceptual (see, e.g., Ref.~\cite{Polyzou:2023vjj}) and practical challenges. Technically, quantization in an explicit frame sacrifices manifest covariance, which, as expected from general symmetry principles, complicates calculations. These wave functions are computed using time-ordered, or ``old-fashioned,'' perturbation theory, which is more involved and less efficient than covariant amplitude calculations. Beyond tree level, calculations can quickly become extremely demanding, as evidenced especially by recent literature (see examples of such calculations in Refs.~\cite{Mueller:2012bn,Beuf:2016wdz,Lappi:2016oup,Beuf:2017bpd,Taels:2022tza,Taels:2023czt,Beuf:2024msh}). In gauge theories, the situation is further complicated by the non-covariant gauge typically chosen, which introduces spurious singularities that require careful regularization and eventually cancel among different terms, often in a highly non-trivial manner.

In this work, we propose a practical method to extract light-cone wave functions from covariant amplitudes. Focusing on \(1 \to 2\) wave functions in a cubic scalar theory, we conjecture a formula that converts covariant off-shell amplitudes into light-cone wave functions. We verify this formula by comparing it with direct light-cone perturbation theory (LCPT) calculations for a few simple yet non-trivial one-loop contributions. Using this formula, we derive compact expressions for \(1 \to 2\) wave functions from known covariant amplitudes, significantly reducing the computational effort compared to direct LCPT methods. We confirm that our results match those obtained recently using the latter approach~\cite{Munier:2025qyz}.

Our paper is organized as follows. Section~\ref{sec:cov-to-LC} reviews the essential building blocks of the covariant and light-cone formalisms for calculating amplitudes and wave functions, respectively, and concludes with a formula expressing light-cone wave functions in terms of covariant amplitudes. In Sec.~\ref{sec:one-loop}, we verify and apply this formula to one-loop calculations. Section~\ref{sec:two-loops} demonstrates how two-loop contributions can be elegantly derived by combining covariant amplitudes. We conclude in Sec.~\ref{sec:conclusion} with a summary and an outline of future directions. Technical details are provided in the appendix.


\section{From covariant amplitudes to light-cone wave functions}
\label{sec:cov-to-LC}

\subsection{Covariant amplitudes}

Throughout, we will consider a cubic massive scalar theory, defined by the Lagrangian density
\begin{equation}
{\cal L}(\varphi,\partial_\mu\varphi) = \frac{1}{2}\partial_\mu\varphi\partial^\mu\varphi - \frac{m^2}{2}\varphi^2 - \frac{\lambda}{3!}\varphi^3.
\end{equation}


\subsubsection{Renormalized Lagrangian and Feynman rules}

Instead of the bare parameters $m$ and $\lambda$, we shall use a renormalized mass $m_R$ and coupling $\lambda_R$, related to the bare quantities $m$ and $\lambda$ through $m^2 \equiv Z_m m_R^2$ and $\lambda \equiv Z_\lambda \lambda_R$, where the renormalization constants $Z_m$ and $Z_\lambda$ depend on $m_R$ and $\lambda_R$. In terms of these quantities, the Lagrangian density reads
\begin{equation}
{\cal L} = {\cal L}_{0R} + {\cal L}_{1R}^{\text{bare}} + {\cal L}_{1R}^{\text{ct}},
\end{equation}
where the free, bare interaction, and counter-term parts are respectively given by
\begin{align}
\begin{aligned}
{\cal L}_{0R} &= \frac{1}{2}\partial_\mu\varphi\partial^\mu\varphi - \frac{m_R^2}{2}\varphi^2, \\
{\cal L}_{1R}^{\text{bare}} &= -\frac{\lambda_R}{3!}\varphi^3, \\
{\cal L}_{1R}^{\text{ct}} &= -(Z_m - 1)\frac{m_R^2}{2}\varphi^2 - (Z_\lambda - 1)\frac{\lambda_R}{3!}\varphi^3.
\end{aligned}
\end{align}
We will work in general space-time dimension $d$, and thus the coupling is dimensionful except for $d = 6$. We introduce the arbitrary momentum scale $\mu$ and define the dimensionless renormalized coupling $\bar{\lambda}_R \equiv \lambda_R \times \mu^{d/2 - 3}$.

Note that we choose not to renormalize the field, meaning that we will need to keep the space-time dimension $d$ as a parameter since we will be left with ultraviolet divergences. We will eventually expand near the dimension in which the theory is perturbatively renormalizable, i.e., $d = 6$. It will prove convenient to define $\delta_6 \equiv 3 - d/6$ in such a way that $\delta_6$ is a small parameter near that critical dimension.

In covariant quantization, after subtracting vacuum fluctuations, the amplitude for the transition from an initial state $i$ to a final state $f$, involving \(n_i\) incoming particles with total momentum \(p_i\) and \(n_f\) outgoing particles with total momentum \(p_f\), is given by the sum of connected diagrams with \(n_i + n_f\) external lines. These diagrams include propagators, cubic interaction vertices, and the necessary two- and three-point counter-term vertices. The corresponding Feynman rules, used to compute the contribution of each diagram to the \(S\)-matrix elements,
\be
S_{fi} = \delta_{fi} + (2\pi)^d \delta^d(p_i - p_f)\, i \mathcal{M}_{i \to f},
\ee
are as follows \cite{Peskin:1995ev}:
\begin{itemize}
    \item For each propagator between two vertices carrying momentum \(p\), assign a factor \(i/(p^2 - m_R^2 + i0^+)\), where \(0^+\) represents a small non-negative real parameter, eventually taken to zero.
    \item For each cubic vertex merging particles carrying the \(d\)-momenta \(p_1, p_2, p_3\), all pointing towards the vertex, assign a factor \((-i\lambda_R)(2\pi)^d \delta^d(p_1 + p_2 + p_3)\).
    \item For each two-point counter-term vertex with legs carrying inward \(d\)-momenta \(p_1\) and \(p_2\), assign a factor \((-im_R^2)(Z_m - 1)(2\pi)^d \delta^d(p_1 + p_2)\).
    \item For each three-point counter-term vertex, assign a factor \((-i\lambda_R)(Z_\lambda - 1)(2\pi)^d \delta^d(p_1 + p_2 + p_3)\).
    \item Integrate over all internal momenta \(p\) with the measure \(d^d p / (2\pi)^d\).
    \item Include an overall combinatorial factor when required by diagram symmetries.
\end{itemize}
To obtain a contribution to the \(S\)-matrix, an appropriate factor must be associated with each external particle, as prescribed by the Lehmann–Symanzik–Zimmermann (LSZ) rules (see, e.g., \cite{Peskin:1995ev,Sterman:1993hfp}). However, since our focus will be on amputated off-shell amplitudes, external leg factors are omitted. Thus, these factors need not be specified here.

The calculation of amplitudes can be efficiently organized using dressed propagators and vertices as building blocks. We now discuss these components and determine the renormalization constants at one-loop accuracy.


\subsubsection{Building blocks of covariant amplitudes}
\label{sec:building-blocks}
\paragraph{Full propagator and mass renormalization}
The first building block of covariant amplitudes is the full propagator. We construct it from the amputated one-particle-irreducible (1-PI) two-point function of momentum \(k\), namely the self-energy, for which we adopt the established notation \(-i\Sigma_R(k^2)\). The dressed propagator of a particle with momentum \(k\) is given by
\begin{equation}
\frac{i}{k^2 - \Sigma_R(k^2) - m_R^2 + i0^+}.
\label{eq:dressed-propagator-0}
\end{equation}
We pick an on-shell renormalization scheme for the mass: the renormalized mass $m_R$ is set to the physical mass, and hence the mass renormalization constant is chosen in such a way that \(\Sigma_R(m_R^2) \equiv 0\). Near the mass shell, the dressed propagator in Eq.~\eqref{eq:dressed-propagator-0} behaves as
\be
\frac{i}{k^2 - \Sigma_R(k^2) - m_R^2 + i0^+} \underset{k^2 \to m_R^2}{\sim} \frac{iZ_\varphi}{k^2 - m_R^2 + i0^+},
\label{eq:dressed-prop-near-massshell}
\ee
{where}
\be
Z_\varphi \equiv \frac{1}{1 - \Sigma'_R(m_R^2)}
\ee
denotes the probability to find the dressed scalar particle in its bare state.

\begin{figure}
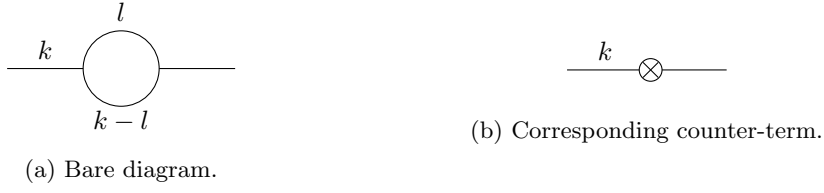

    \begin{center}
    \begin{subfigure}[t]{0.4\textwidth}
    \centerline{\adjustbox{valign=c}{\onetooneoneloop}}
    \caption{Bare diagram.}
    \label{subfig:self-energy-one-loop-bare}
    \end{subfigure}
    \begin{subfigure}[t]{0.4\textwidth}
    \centerline{\adjustbox{valign=c}{\onetooneoneloopct{${k}$}{}{}}}
    \caption{Corresponding counter-term.}
    \label{subfig:self-energy-one-loop-counterterm}
    \end{subfigure}
    \end{center}
    \caption{Order $\bar\lambda_R^2$ contributions to the self-energy $-i\Sigma_R(k^2)$.}
    \label{fig:self-energy-one-loop}
\end{figure}

Let us expand $\Sigma_R$ to one-loop accuracy. At this order, the unrenormalized self-energy, evaluated from the Lagrangian without counter-terms, \({\cal L}_{0R} + {\cal L}_{1R}^{\text{bare}}\), stems from a single diagram; see Fig.~\ref{subfig:self-energy-one-loop-bare}. Its expression reads
\begin{equation}
\left.-i\Sigma(k^2)\right|_\text{1 loop} = \frac{(-i\lambda_R)^2}{2}\int\frac{d^d l}{(2\pi)^d}\frac{i}{l^2 - m_R^2 + i0^+}\frac{i}{(k-l)^2 - m_R^2 + i0^+}.
\label{eq:Sigma-from-diagram}
\end{equation}
The evaluation of \(\Sigma(k^2)\) is a textbook calculation (see e.g. Ref.~\cite{Sterman:1993hfp}):
\begin{equation}
\left.\Sigma(k^2)\right|_\text{1 loop} = -\frac{\lambda_R^2}{2}\frac{m_R^{2-2\delta_6}}{(4\pi)^{3-\delta_6}}\Gamma(\delta_6-1)\int_0^1 dx\left(1 - x(1-x)\frac{k^2}{m_R^2} - i0^+\right)^{1-\delta_6}.
\end{equation}
Keeping only the divergent and constant terms in the limit \(\delta_6 \to 0\):
\begin{equation}
\left.\Sigma(k^2)\right|_\text{1 loop} = -\frac{\bar\lambda_R^2}{2}\frac{m_R^2}{(4\pi)^3}\left[\left(\frac{1}{6}\frac{k^2}{m_R^2} - 1\right)\left(\frac{1}{\delta_6} - \gamma_E + \ln 4\pi + 1 + \ln\frac{\mu^2}{m_R^2}\right) - {\cal J}\left(\frac{k^2}{m_R^2} + i0^+\right)\right],
\end{equation}
where
\begin{equation}
{\cal J}(x) \equiv \frac{5x}{18} - \frac{4}{3} + \frac{x}{3}\times\begin{cases}\left(1 - \frac{4}{x}\right)^{3/2}\atanh\frac{1}{\sqrt{1 - {4}/{x}}}&\text{for $x<0$}\,,\\
\left(\frac{4}{x}-1\right)^{3/2}\atan\frac{1}{\sqrt{{4}/{x}-1}}&\text{for $0<x<4$}\,.
\end{cases}
\label{eq:defJ}
\end{equation}
We adopt this expression, up to a trivial continuation at \(x = 0\), as the definition of the function on the cut complex plane \(\mathbb{C} \setminus [4, +\infty)\).

The one-loop self-energy, after including the counter-term diagram shown in Fig.~\ref{subfig:self-energy-one-loop-counterterm} and enforcing the on-shell scheme, is given by
\begin{equation}
\begin{aligned}
\left.\Sigma_R(k^2)\right|_\text{1 loop}&=\left.\Sigma(k^2)\right|_\text{1 loop} -\left.\Sigma(m_R^2)\right|_\text{1 loop}\\ 
&= -\frac{\bar\lambda_R^2}{2}\frac{m_R^2}{(4\pi)^3}\left[\frac{1}{6}\frac{k^2 - m_R^2}{m_R^2}\left(\frac{1}{\delta_6} - \gamma_E + \ln 4\pi + 1 + \ln\frac{\mu^2}{m_R^2}\right) + {\cal J}(1) - {\cal J}\left(\frac{k^2}{m_R^2} + i0^+\right)\right].
\end{aligned}
\label{eq:SigmaR-one-loop}
\end{equation}
The constant \({\cal J}(1)\) reads
\begin{equation}
{\cal J}(1)= \frac{1}{6}\left(\sqrt{3}\pi - \frac{19}{3}\right).
\end{equation}

The residual divergence in Eq.~(\ref{eq:SigmaR-one-loop}) as \(\delta_6 \to 0\) would be absorbed by renormalizing the field \(\varphi\), but as already mentioned, we choose not to renormalize it and instead keep the dimension of space-time slightly off \(d=6\). We may however introduce the constant
\be
\left.Z_\varphi^{\MSbar}\right|_\text{1 loop}\equiv 1-\frac{\bar\lambda_R^2}{12(4\pi)^3}\left(\frac{1}{\delta_6}-\gamma_E+\ln 4\pi\right)
\ee
in terms of which
\begin{multline}
\left.\Sigma_R(k^2)\right|_\text{1 loop} =\left(\left.Z_\varphi^{\MSbar}\right|_\text{1 loop}-1\right)(k^2-m_R^2)\\ -\frac{\bar\lambda_R^2}{2}\frac{m_R^2}{(4\pi)^3}\left[\frac{1}{6}\frac{k^2 - m_R^2}{m_R^2}\left(1 + \ln\frac{\mu^2}{m_R^2}\right) + {\cal J}(1) - {\cal J}\left(\frac{k^2}{m_R^2} + i0^+\right)\right].
\label{eq:SigmaR-one-loop-rewritten}
\end{multline}

Let us highlight a couple of properties of the expression of the self-energy \(\Sigma_R(k^2)\), that will be useful in the following discussion:
\begin{itemize}
    \item \(\Sigma_R(k^2)\) develops an imaginary part above the two-particle production threshold, specifically for \(k^2 > 4m_R^2\). At one-loop order, this imaginary part arises from the function \(\mathcal{J}(x)\) in Eq.~\eqref{eq:SigmaR-one-loop-rewritten}, which exhibits a logarithmic branch cut along the real axis for \(x > 4\). The discontinuity across this cut is given by
\be
\text{Disc}_{x}\mathcal{J}(x) \equiv \lim_{\epsilon \to 0^+}\left[\mathcal{J}(x+i\epsilon) - \mathcal{J}(x-i\epsilon)\right] = \frac{x}{3}\left(1 - \frac{4}{x}\right)^{3/2} \times (-i\pi)\times\mathbbm{1}_{\{x > 4\}}.
\ee
This, in turn, translates into the following discontinuity for \(\Sigma_R(k^2)\):
\be
\text{Disc}_{k^2}\left.\Sigma_R(k^2)\right|_{\text{1 loop}} = -i\frac{2\bar{\lambda}_R^2}{3(16\pi)^2}k^2\left(1 - \frac{4m_R^2}{k^2}\right)^{3/2}\mathbbm{1}_{\{k^2 > 4m_R^2\}}=2i\Im\left.\Sigma_R(k^2)\right|_{\text{1 loop}}.
\label{eq:Sigma-one-loop-disc}
\ee

\item For any \(k^2\) within its domain of analyticity, \(\Sigma_R(k^2)\) can be expressed using a dispersion relation subtracted at a point \(s_0 < 4m_R^2\):
\begin{equation}
\Sigma_R(k^2) = \Sigma_R(s_0) + (k^2 - s_0) \int_{4m_R^2}^{+\infty} \frac{ds}{2i\pi} \frac{\text{Disc}_{s}\Sigma_R(s)}{(s - s_0)(s - k^2)}.
\label{eq:dispersion-Sigma-0}
\end{equation}
Setting the subtraction point \(s_0\) to \(m_R^2\) causes the first term to vanish in the renormalization scheme defined above. By differentiating the previous formula with respect to \(s_0\), setting subsequently \(s_0 = m_R^2\), and combining the result with Eq.~\eqref{eq:dispersion-Sigma-0}, we obtain the relation
\begin{equation}
\frac{\Sigma_R(k^2)}{(k^2 - m_R^2)^2} = \frac{\Sigma'_R(m_R^2)}{k^2 - m_R^2} + \int_{4m_R^2}^{+\infty} \frac{ds}{2i\pi} \frac{\text{Disc}_{s}\Sigma_R(s)}{(s - m_R^2)^2(s - k^2)}.
\label{eq:dispersion-Sigma}
\end{equation}
\end{itemize}


\paragraph{Vertex corrections and coupling renormalization}

\begin{figure}
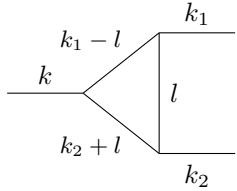
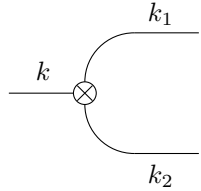

    \begin{center}
    \begin{subfigure}[t]{0.4\textwidth}
    \centerline{\adjustbox{valign=c}{\triangle}}
    \caption{Bare one-loop diagram.}
    \label{subfig:triangle-one-loop-bare}
    \end{subfigure}
    \begin{subfigure}[t]{0.4\textwidth}
    \centerline{\adjustbox{valign=c}{\trianglect}}
    \caption{Corresponding counter-term.}
    \label{subfig:triangle-one-loop-counterterm}
    \end{subfigure}
    \end{center}
    \caption{Order $\bar\lambda_R^3$ contributions to the vertex function $i\Gamma_{3R}$.}
    \label{fig:triangle-one-loop}
\end{figure}

The second building block of covariant amplitudes is the amputated one-particle-irreducible three-point function.

Up to one-loop accuracy, the contribution computed from the Lagrangian without counter-terms  is given by\footnote{Let us clarify our notation. The subscript ``1 loop'' will consistently denote ``up to one-loop accuracy.'' When referring to a term of a given order \(n\), we will use the subscript \(\mathcal{O}(\bar\lambda_R^n)\).}
\be
\left.i\Gamma_3(k^2, k_1^2, k_2^2)\right|_{\text{1 loop}}=-i\lambda_R+\left.i\Gamma_3(k^2, k_1^2, k_2^2)\right|_{{\cal O}(\bar\lambda_R^3)},
\ee
where the second term is the expression of the diagram displayed in Fig.~\ref{subfig:triangle-one-loop-bare}:
\begin{equation}
\left.i\Gamma_3(k^2, k_1^2, k_2^2)\right|_{{\cal O}(\bar\lambda_R^3)} = (-i\lambda_R)^3 \int \frac{d^d l}{(2\pi)^d} \frac{i}{l^2 - m_R^2 + i0^+} \frac{i}{(k_1 - l)^2 - m_R^2 + i0^+} \frac{i}{(k_2 + l)^2 - m_R^2 + i0^+}.
\label{eq:Gamma3-from-diagram}
\end{equation}
Using standard techniques~\cite{Sterman:1993hfp}, this $d$-dimensional integral can be rewritten as an integral over two Feynman parameters:
\begin{equation}
\left.\Gamma_3(k^2, k_1^2, k_2^2)\right|_{{\cal O}(\bar\lambda_R^3)} = -\frac{\lambda_R^3}{(4\pi)^{3 - \delta_6}} \Gamma(\delta_6) \int_0^1 du \int_0^{1 - u} dv \left[m_R^2 - (1 - u - v)(uk_1^2 + vk_2^2) - uv k^2 - i0^+\right]^{-\delta_6}.
\end{equation}
The ultraviolet divergence arising as $\delta_6 \to 0$ is fully removed by introducing the appropriate counter-term (see Fig.~\ref{subfig:triangle-one-loop-counterterm}), yielding the renormalized vertex function
\begin{equation}
\left.\Gamma_{3R}(k^2, k_1^2, k_2^2)\right|_{{\cal O}(\bar\lambda_R^3)} = \left.\Gamma_3(k^2, k_1^2, k_2^2)\right|_{{\cal O}(\bar\lambda_R^3)} - \lambda_R\left( \left.Z_\lambda\right|_{\text{1 loop}}-1\right).
\label{eq:def-Gamma3R}
\end{equation}
The simplest renormalization scheme for our purpose is $\overline{\text{MS}}$. In that scheme, the one-loop coupling renormalization constant takes the form
\begin{equation}
\left.Z_\lambda\right|_{\text{1 loop}} = 1 - \frac{\bar\lambda_R^2}{2(4\pi)^3} \left(\frac{1}{\delta_6} - \gamma_E + \ln 4\pi\right).
\end{equation}
The renormalized one-loop vertex function, retaining only the finite terms in the $\delta_6 \to 0$ limit, reads
\begin{equation}
\left.\Gamma_{3R}(k^2, k_1^2, k_2^2)\right|_{{\cal O}(\bar\lambda_R^3)} = \frac{\bar\lambda_R^3}{(4\pi)^3} \int_0^1 du \int_0^{1 - u} dv \ln\left[\frac{m_R^2 - (1 - u - v)(uk_1^2 + vk_2^2) - uv k^2 - i0^+}{\mu^2}\right].
\label{eq:Gamma3R}
\end{equation}

Expressions for \(\Gamma_3\) in arbitrary space-time dimensions exist, though they involve complicated special functions (see Ref.~\cite{tHooft:1978jhc,Fleischer:2003rm}). \(\Gamma_3\) exhibits discontinuities, with explicit expressions provided in Ref.~\cite{Abreu:2015zaa}, specifically in Eq.~(D.28) for masses all set to \(m_R\) and \(d = 6\).\footnote{See also Ref.~\cite{Muhlbauer:2022ylo}, Eq.~(246). Note, however, that this expression differs from the equivalent result in Ref.~\cite{Abreu:2015zaa} when the dimension is not 4. We attribute this discrepancy to a non-standard treatment of angular integration for arbitrary space-time dimensions in Ref.~\cite{Muhlbauer:2022ylo}.}

In the massless limit \(m_R \to 0\), however, simple closed-form expressions can be derived. The relevant cases for the present discussion are detailed in Appendix~\ref{sec:appendix-triangle}.


\subsection{Light-cone wave functions}

We now turn to a Hamiltonian formulation of the theory in a light-cone frame. We begin by defining the light-cone coordinates. Starting from a Lorentz frame where a generic \(d\)-vector \(v\) has coordinates \((v^0, v^\perp, v^{d-1})\), with \(v^\perp \equiv (v^1, \cdots, v^{d-2})\), the light-cone coordinates of the same vector are defined as $(v^+,v^-,v^\perp)$, where
\begin{equation}
v^\pm \equiv \frac{v^0 \pm v^{d-1}}{\sqrt{2}}.
\end{equation}
The ``$+$'' component of a $d$-position vector $(x^+,x^-,x^\perp)$ is defined to be the light-cone time, and the remaining components will be grouped in the ``spatial'' vector $\vec x\equiv(x^-,x^\perp)$. The ``$-$'' component of a $d$-momentum vector $(k^+,k^-,k^\perp)$ serves as our energy variable. We denote by \(E_R(\vec{k})\) the light-cone energy $k^-$ of an on-shell particle with mass \(m_R\) and momentum \(\vec{k} \equiv (k^+, k^\perp)\). It is given by
\begin{equation}
E_R(\vec{k}) = \frac{k^{\perp 2} + m_R^2}{2k^+}.
\end{equation}
For later convenience, we introduce the following notations:
\begin{equation}
E_R(\vec{k}_1, \vec{k}_2, \cdots, \vec{k}_n) \equiv \sum_{i=1}^n E_R(\vec{k}_i),
\quad
\Delta_R(\vec{k}_1, \vec{k}_2, \cdots, \vec{k}_n) \equiv E_R\left(\sum_{i=1}^n \vec{k}_i\right) - E_R(\vec{k}_1, \vec{k}_2, \cdots, \vec{k}_n).
\end{equation}
The on-shell integration measure for a particle with momentum \(\vec{k}\) is defined as
\begin{equation}
\widetilde{dk} = \frac{dk^+}{2k^+} \mathbbm{1}_{\{k^+ \geq 0\}} d^{d-2}k^\perp.
\end{equation}
Additionally, we associate with any \(d\)-momentum \(k\) a modified \(d\)-momentum \(\tilde{k}\), which shares the same spatial components \(\vec{\tilde{k}} \equiv \vec{k}\) but has an energy component given by \(\tilde{k}^- \equiv E_R(\vec{k})\).

The Hamiltonian governing time evolution in this light-cone frame is given by \(\mathcal{H} = \mathcal{H}_{0R} + \mathcal{H}_{1R}^\text{bare} + \mathcal{H}_{1R}^\text{ct}\), where
\be
\begin{aligned}
\mathcal{H}_{0R} &= \int d^{d-1}\vec{x} \left(\frac{1}{2}(\partial^\perp \varphi)^2 + \frac{m_R^2}{2} \varphi^2\right), \\
\mathcal{H}_{1R}^\text{bare} &= \int d^{d-1}\vec{x} \,\frac{\lambda_R}{3!} \varphi^3, \\
\mathcal{H}_{1R}^\text{ct} &= \int d^{d-1}\vec{x} \left(\left(Z_m - 1\right) \frac{m_R^2}{2} \varphi^2 + \left(Z_\lambda - 1\right) \frac{\lambda_R}{3!} \varphi^3\right).
\end{aligned}
\label{eq:H-phi}
\ee


\subsubsection{Perturbation theory for wave functions}

We shall work with the usual basis for the space of the states of a free particle as the eigenstates of ${\mathcal{H}}_{0R}$, uniquely labeled by their momenta \(\vec{k}\). We adopt the following normalization convention:
\begin{equation}
\braket{\vec{k}'}{\vec{k}} = 2k^+ (2\pi)^{d-1} \delta^{d-1}(\vec{k}' - \vec{k}).
\end{equation}
The canonical basis of the Fock space of multi-particle states consists of symmetrized tensor products of these states, \(\ket{\vec{k}_1, \cdots, \vec{k}_n}\). We set the normalizations in such a way that the completeness relation reads
\begin{equation}
\sum_{n} \int \widetilde{dk_1} \cdots \widetilde{dk_n} \, \frac{1}{n!} \ket{\vec{k}_1, \cdots, \vec{k}_n} \bra{\vec{k}_1, \cdots, \vec{k}_n} = \mathbbm{1}.
\end{equation}
We denote a generic basis vector by \(\ket{\Phi}\) and its number of particles by \(N_\Phi\). In these notations,
\begin{equation}
\sum_{\ket{\Phi}} \frac{1}{N_\Phi!} \ket{\Phi} \bra{\Phi} = \mathbbm{1},
\end{equation}
where the sum over \(\ket{\Phi}\) includes integration over the on-shell momenta of its $N_\Phi$ constituent particles. The energy of a Fock state \(\ket{\Phi}\) is given by the sum of the energies of its constituent particles, denoted by \(E_R(\Phi)\).

The probability amplitude for a system initially prepared as a single scalar particle in the normalized state
\begin{equation}
\ket{\phi} \equiv \int \widetilde{dk} \, \phi(\vec{k}) \ket{\vec{k}},
\quad \text{with} \quad
\int \widetilde{dk} \, \left|\phi(\vec{k})\right|^2 = 1,
\label{eq:wavepacket}
\end{equation}
to be observed in a state of \(n > 0\) free particles with definite momenta \(\vec{k}_1, \cdots, \vec{k}_n\), namely, its \(1 \to n\) light-cone wave function, can be computed using the following expression~\cite{Bjorken:1970ah,Lepage:1980fj}:
\begin{mdframed}[innertopmargin=-5pt, innerbottommargin=5pt, leftmargin=0pt, rightmargin=0pt,skipabove=2pt,skipbelow=2pt]
\begin{multline}
\psi_{\phi \to n\times\varphi}(\vec{k}_1, \cdots, \vec{k}_n) = \int \widetilde{dk} \, \frac{\sqrt{Z_\varphi} \, \phi(\vec{k})}{\tilde{k}^- - E_R(\vec{k}_1, \cdots, \vec{k}_n)} \Bigg( \mel{\vec{k}_1, \cdots, \vec{k}_n}{\mathcal{H}_{1R}}{\vec{k}} \\
+ \sum_{j=1}^{+\infty} \sum_{\ket{\Phi_j}, \cdots, \ket{\Phi_1}\neq \ket{\vec{k}}} \frac{1}{N_{\Phi_j}! \cdots N_{\Phi_1}!} \frac{\mel{\vec{k}_1, \cdots, \vec{k}_n}{\mathcal{H}_{1R}}{\Phi_j} \cdots \mel{\Phi_1}{\mathcal{H}_{1R}}{\vec{k}}}{\left(\tilde{k}^- - E_R(\Phi_j)\right) \cdots \left(\tilde{k}^- - E_R(\Phi_1)\right)} \Bigg).
\label{eq:LCPT}
\end{multline}
\end{mdframed}
The matrix elements of the Hamiltonian \(\mathcal{H}_{1R}\) in Eq.~\eqref{eq:H-phi} can be reduced to products of elementary matrix elements between one- and two-particle states. The latter are given by
\begin{equation}
\begin{aligned}
\mel{\vec{k}_1, \vec{k}_2}{\mathcal{H}_{1R}^\text{bare}}{\vec{k}} &= \lambda_R (2\pi)^{d-1} \delta^{d-1}(\vec{k} - \vec{k}_1 - \vec{k}_2), \\
\mel{\vec{k}_1,\vec{k}_2}{\mathcal{H}_{1R}^\text{ct}}{\vec{k}} &= \lambda_R (Z_\lambda - 1) (2\pi)^{d-1} \delta^{d-1}(\vec{k} - \vec{k}_1 - \vec{k}_2), \\
\mel{\vec{k}}{\mathcal{H}_{1R}}{\vec{k}_1, \vec{k}_2} &= \mel{\vec{k}_1, \vec{k}_2}{\mathcal{H}_{1R}}{\vec{k}}, \\
\mel{\vec{k}'}{\mathcal{H}_{1R}^\text{ct}}{\vec{k}} &= m_R^2 (Z_m - 1) (2\pi)^{d-1} \delta^{d-1}(\vec{k} - \vec{k}').
\end{aligned}
\label{eq:matrix-elements-phi3}
\end{equation}
Since all elementary interactions conserve the \((d-1)\)-momentum, unfolding the matrix elements of the Hamiltonian in Eq.~\eqref{eq:LCPT} reveals that the momentum \(\vec{k}\) coincides with \(\sum_{i=1}^n\vec{k}_i\). Consequently, the so-called ``energy denominators'' will eventually reduce as 
\be
\begin{aligned}
\tilde{k}^- - E_R(\vec{k}_1, \cdots, \vec{k}_n)&\to\Delta_R(\vec{k}_1, \cdots, \vec{k}_n)\\
\left(\tilde{k}^- - E_R(\Phi_j)\right) \cdots \left(\tilde{k}^- - E_R(\Phi_1)\right)&\to \Delta_R(\Phi_j) \cdots \Delta_R(\Phi_1).
\end{aligned}
\ee

In this perturbative framework, the normalization constant \(Z_\varphi\) is determined once all other terms have been evaluated, via the unitarity condition. The latter translates into the following constraint:
\begin{equation}
\sum_{n} \frac{1}{n!} \int \widetilde{dk_1} \cdots \widetilde{dk_n} \, \left|\psi_{\phi \to n\times\varphi}(\vec{k}_1, \cdots, \vec{k}_n)\right|^2 = 1.
\end{equation}
Note that \(Z_\varphi\) could be computed purely diagrammatically, as detailed in Ref.~\cite{Munier:2025qyz}. However, for the present work, the ``standard'' formulation described therein, where the perturbative series takes the widely-used form of Eq.~\eqref{eq:LCPT}, is the most suitable.

For the mass and coupling renormalization constants, we adopt the values obtained from the covariant approach, in accordance with the renormalization conditions discussed earlier.


\subsubsection{Light-cone wave functions from covariant amplitudes}

We specialize to amplitudes describing the decay of a single initial particle into two particles with momenta \(\vec{k}_1\) and \(\vec{k}_2\), where \(k_1^+\) and \(k_2^+\) are positive. Our goal is to extract the light-cone wave function for a particle initially prepared in the state \(\ket{\phi}\) to be found in the two-particle state \(\ket{\vec{k}_1, \vec{k}_2}\), from a covariant amplitude constructed from the building blocks of Sec.~\ref{sec:building-blocks}.

We conjecture the following relationship between these two objects:
\begin{mdframed}[innertopmargin=-5pt, innerbottommargin=5pt, leftmargin=0pt, rightmargin=0pt,skipabove=2pt,skipbelow=2pt]
\begin{multline}
\psi_{\phi\to\varphi\varphi}(\vec{k}_1, \vec{k}_2) = \int \frac{d^d k}{(2\pi)^d} \sqrt{Z_\varphi} \, \phi(\vec{k}) \int_{-\infty}^{+\infty} \frac{dk_1^-}{2\pi} \, 2k_1^+ \int_{-\infty}^{+\infty} \frac{dk_2^-}{2\pi} \, 2k_2^+ \\
\times \frac{(2\pi)^d \delta^d(k - k_1 - k_2) \, \Gamma_{3R}(k^2, k_1^2, k_2^2)}{\left(m_R^2 - k^2 + i0^+\right) \left[k_1^2 - \Sigma_R(k_1^2) - m_R^2 + i0^+\right] \left[k_2^2 - \Sigma_R(k_2^2) - m_R^2 + i0^+\right]},
\label{eq:relation-scalar}
\end{multline}
\end{mdframed}
where \(\Sigma_R\) and \(\Gamma_{3R}\) are the 1-PI amputated two- and three-point functions introduced in Sec.~\ref{sec:building-blocks}. In this formula, the integration variables are kept real, and the singularities are regularized by explicitly retaining the small imaginary parts \(i0^+\) in all factors, including \(\Gamma_{3R}\) and \(\Sigma_R\). 

The integrand differs from a regular three-point function. Indeed, while the outgoing particles are represented by full propagators, a bare propagator is used for the incoming particle. Furthermore, the Feynman prescription \(i0^+\) has an unconventional sign for the latter, causing the integrand to differ from that of a regular three-point function.

The integral over \(k\) can be performed using the Dirac delta function, which identifies \(k\) with \(k_1 + k_2\). We then express the momenta in the denominators in terms of their light-cone components, using the kinematical relations
\begin{equation}
\begin{aligned}
k_{i}^2 - m_R^2 &= 2k_{i}^+ \left(k_{i}^- - \tilde{k}_{i}^-\right), \quad \text{for } i = 1, 2, \\
m_R^2 - (k_1 + k_2)^2 &= 2(k_1^+ + k_2^+) \left[E_R(\vec{k}_1 + \vec{k}_2) - k_1^- - k_2^-\right].
\end{aligned}
\label{eq:kinematics-cov-to-LC}
\end{equation}
The formula for the wave function simplifies to
\begin{multline}
\psi_{\phi\to\varphi\varphi}(\vec{k}_1, \vec{k}_2) = \frac{\phi(\vec{k}_1 + \vec{k}_2)}{2(k_1^+ + k_2^+)} \int_{-\infty}^{+\infty} \frac{dk_1^-}{2\pi} \frac{dk_2^-}{2\pi} \frac{1}{E_R(\vec{k}_1 + \vec{k}_2) - k_1^- - k_2^- + i0^+} \\
\times \frac{\Gamma_{3R}\left((k_1 + k_2)^2, k_1^2, k_2^2\right)}{\left(k_1^- - \tilde{k}_1^- - \frac{\Sigma_R(k_1^2)}{2k_1^+} + i0^+\right) \left(k_2^- - \tilde{k}_2^- - \frac{\Sigma_R(k_2^2)}{2k_2^+} + i0^+\right)},
\label{eq:relation-scalar-integrated}
\end{multline}
where \(k_1^2\), \(k_2^2\), and \((k_1 + k_2)^2\) are now understood to be given by Eq.~\eqref{eq:kinematics-cov-to-LC}.

Next, we complexify the integration variable \(k_2^-\). The integrand exhibits a single singularity in the upper-half plane: a simple pole at \(k_2^- = E_R(\vec{k}_1 + \vec{k}_2) - k_1^- + i0^+.\) All other singularities, including poles and branch cuts, are confined to the lower-half plane. By closing the contour with a semicircular arc at infinity in the upper-half plane, where the integrand vanishes as $\left(1/k_1^-\right)^2$, we turn the linear integration path into a closed loop. Cauchy’s theorem then allows us to evaluate the integral by computing the residue at the pole. This reduces the expression to a single integral over a real variable, involving covariant two- and three-point functions:
\begin{mdframed}[innertopmargin=-5pt, innerbottommargin=5pt, leftmargin=0pt, rightmargin=0pt,skipabove=2pt,skipbelow=2pt]
\begin{equation}
\psi_{\phi\to\varphi\varphi}(\vec{k}_1, \vec{k}_2) = \frac{\phi(\vec{k}_1 + \vec{k}_2)}{2(k_1^+ + k_2^+)} \int_{-\infty}^{+\infty} \frac{dk_1^-}{2i\pi} \frac{\Gamma_{3R}\left(m_R^2, k_1^2, k_2^2\right)}{\left(k_1^- - \tilde{k}_1^- - \frac{\Sigma_R(k_1^2)}{2k_1^+} + i0^+\right) \left(\Delta_R(\vec{k}_1, \vec{k}_2) + \tilde{k}_1^- - k_1^- - \frac{\Sigma_R(k_2^2)}{2k_2^+} + i0^+\right)},
\label{eq:relation-scalar-integrated-integrated}
\end{equation}
\end{mdframed}
where \(k_1^2\) and \(k_2^2\) are given by Eq.~\eqref{eq:kinematics-cov-to-LC}, with \(k_2^-\) replaced by \(E_R(\vec{k}_1 + \vec{k}_2) - k_1^- + i0^+\). This can also be expressed in terms of \(\Delta_R(\vec{k}_1, \vec{k}_2)\). Explicitly,
\be
\begin{aligned}
k_1^2&=2k_1^+(k_1^--\tilde k_1^-)+m_R^2,\\
k_2^2&=2k_2^+\left[\Delta_R(\vec k_1,\vec k_2)+\tilde k_1^--k_1^-\right]+m_R^2+i0^+.
\end{aligned}
\ee

We claim that inserting the expressions for the self-energy \(\Sigma_R\) and the three-point function \(\Gamma_{3R}\) into Eq.~\eqref{eq:relation-scalar-integrated-integrated} and performing the remaining one-dimensional integral yields the desired wave function.

Let us note that the incoming particle propagator's only role is to enforce, via the residue at this pole, that the incoming particle be on its mass shell — a condition equivalently imposed by a delta function. However, keeping explicit the propagator, as in Eq.~\eqref{eq:relation-scalar}, is a more symmetric writing, since the integrand is manifestly a product of fully covariant elements (propagators, self-energies, dressed vertices). It is also a much more convenient starting point for the proof of the term-by-term equivalence with LCPT provided in Sec.~\ref{sec:validation} below.


\paragraph{Tree-level two-particle wave function}

As a first check, let us apply Eq.~(\ref{eq:relation-scalar-integrated-integrated}) to the simplest case by evaluating the two-particle wave function at tree level. Here, we set \(\Gamma_{3R} \to -\lambda_R\), \(\sqrt{Z_\varphi} \to 1\), and disregard the self-energy terms \(\Sigma_R\) in the denominators. After simplification, we find
\begin{equation}
\left.\psi_{\phi\to\varphi\varphi}(\vec{k}_1, \vec{k}_2)\right|_\text{tree} = \frac{\phi(\vec{k}_1 + \vec{k}_2)}{2(k_1^+ + k_2^+)} \int_{-\infty}^{+\infty} \frac{dk_1^-}{2i\pi} \frac{-\lambda_R}{\left(k_1^- - \tilde{k}_1^- + i0^+\right)\left(\Delta_R(\vec{k}_1, \vec{k}_2) + \tilde{k}_1^- - k_1^- + i0^+\right)}.
\end{equation}

To evaluate this integral, we complexify the variable \(k_1^-\). By adding a semicircle at infinity in the lower half-plane, we transform the integral into a closed contour integral. Using Cauchy's theorem, we evaluate the integral by capturing the residue at the pole \(k_1^- = \tilde{k}_1^- - i0^+.\) The resulting tree-level wave function is given by
\begin{equation}
\left.\psi_{\phi\to\varphi\varphi}(\vec{k}_1, \vec{k}_2)\right|_\text{tree} = \frac{\phi(\vec{k}_1 + \vec{k}_2)}{2(k_1^+ + k_2^+)} \frac{\lambda_R}{\Delta_R(\vec{k}_1, \vec{k}_2)}.
\label{eq:scalar-tree}
\end{equation}

This result is in exact agreement with the expression obtained from the rules of LCPT; see Eq.~(120) in Ref.~\cite{Munier:2025qyz}.


\section{One-loop corrections to the two-particle wave function}
\label{sec:one-loop}

\begin{figure}
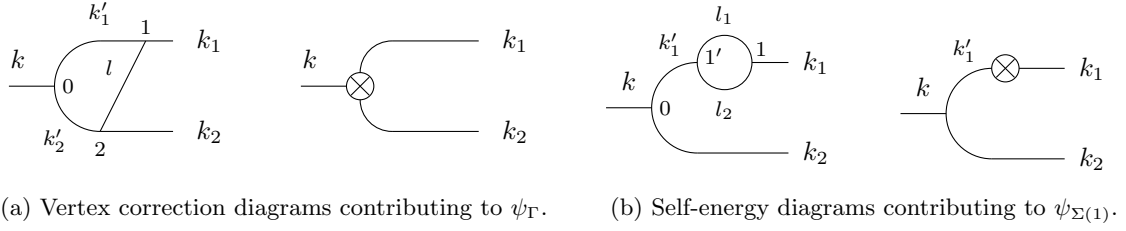

\begin{center}
    \begin{subfigure}[b]{.45\textwidth}
        \centerline{\scalarvertex\quad\quad\scalarvertexct}
        \caption{Vertex correction diagrams contributing to $\psi_{\Gamma}$.}
        \label{subfig:scalarvertex}
    \end{subfigure}
    \begin{subfigure}[b]{0.45\textwidth}
        \centerline{\scalarselfenergyone\quad\quad\scalarselfenergyonect}
        \caption{Self-energy diagrams contributing to $\psi_{\Sigma (1)}$.}
        \label{subfig:scalarselfenergyone}
    \end{subfigure}
    \end{center}
    \caption{A subset of the one-loop diagrams contributing to $\psi_{\phi\to\varphi\varphi}$. We have labeled the propagators by the $d$-dimensional momenta that flow through them, from left to right, and the trivalent vertices by (possibly primed) numbers.}
    \label{fig:scalaroneloop}
\end{figure}

We now expand Eqs.~\eqref{eq:relation-scalar} and~\eqref{eq:relation-scalar-integrated-integrated} perturbatively to order \(\lambda_R^3\). The contribution of that order can be expressed~as
\begin{equation}
\left.\psi_{\phi\to\varphi\varphi}(\vec{k}_1, \vec{k}_2)\right|_{{\cal O}(\bar\lambda_R^3)} = \psi_{\Gamma}(\vec{k}_1, \vec{k}_2) + \psi_{\Sigma (0)}(\vec{k}_1, \vec{k}_2) + \psi_{\Sigma (1)}(\vec{k}_1, \vec{k}_2) + \psi_{\Sigma (2)}(\vec{k}_1, \vec{k}_2),
\label{eq:relation-scalar-1-loop}
\end{equation}
where the first term, \(\psi_{\Gamma}(\vec{k}_1, \vec{k}_2)\), corresponds to the vertex correction diagrams shown in Fig.~\ref{subfig:scalarvertex}, and the other terms to self-energy corrections. The diagrams corresponding to the contribution \(\psi_{\Sigma (1)}\) are shown in Fig.~\ref{subfig:scalarselfenergyone}. We will provide detailed expressions shortly, starting either from Eq.~\eqref{eq:relation-scalar} or from Eq.~\eqref{eq:relation-scalar-integrated-integrated}, depending on the purpose.

Our first objective is to verify our formulas. We will then use the covariant expressions for \(\Gamma_{3R}\) and \(\Sigma_R\) to directly recover the known results for the wave function.

\subsection{Checking the equivalence with LCPT}
\label{sec:validation}

Here, we demonstrate that the right-hand side of Eq.~\eqref{eq:relation-scalar} or~\eqref{eq:relation-scalar-integrated-integrated} can be algebraically manipulated to exactly match the relevant terms in the LCPT series~\eqref{eq:LCPT}. Although this process is extremely tedious, we present the full calculation in detail for the one-loop diagrams addressed in this section. This detailed exposition clarifies that the procedure generalizes to any diagram.

For this purpose, it is convenient to start from the expansion of Eq.~\eqref{eq:relation-scalar}, in which case
\begin{equation}
\psi_{\Gamma}(\vec{k}_1, \vec{k}_2) = \int \frac{d^d k}{(2\pi)^d} \phi(\vec{k}) \int_{-\infty}^{+\infty} \frac{dk_1^-}{2\pi} 2k_1^+ \int_{-\infty}^{+\infty} \frac{dk_2^-}{2\pi} 2k_2^+ \frac{(2\pi)^d \delta^d(k - k_1 - k_2) \left.\Gamma_{3R}(k^2, k_1^2, k_2^2)\right|_{{\cal O}(\bar\lambda_R^3)}}{\left(m_R^2 - k^2 + i0^+\right) \left(k_1^2 - m_R^2 + i0^+\right) \left(k_2^2 - m_R^2 + i0^+\right)}.
\label{eq:psi-Gamma3}
\end{equation}
The remaining terms correspond to self-energy corrections. The first of these terms is given by
\begin{equation}
\psi_{\Sigma (0)}(\vec{k}_1, \vec{k}_2) = \int \frac{d^d k}{(2\pi)^d} \phi(\vec{k}) \int_{-\infty}^{+\infty} \frac{dk_1^-}{2\pi} 2k_1^+ \int_{-\infty}^{+\infty} \frac{dk_2^-}{2\pi} 2k_2^+ \frac{-\lambda_R \left(\left.\sqrt{Z_\varphi}\right|_{\text{1 loop}} - 1\right) (2\pi)^d \delta^d(k - k_1 - k_2)}{\left(m_R^2 - k^2 + i0^+\right) \left(k_1^2 - m_R^2 + i0^+\right) \left(k_2^2 - m_R^2 + i0^+\right)}.
\label{eq:psi-Sigma0}
\end{equation}
The subsequent terms are given by
\begin{equation}
\begin{aligned}
\psi_{\Sigma (1)}(\vec{k}_1, \vec{k}_2) &= \int \frac{d^d k}{(2\pi)^d} \phi(\vec{k}) \int_{-\infty}^{+\infty} \frac{dk_1^-}{2\pi} 2k_1^+ \int_{-\infty}^{+\infty} \frac{dk_2^-}{2\pi} 2k_2^+ \frac{-\lambda_R \left.\Sigma_R(k_1^2)\right|_{\text{1 loop}} (2\pi)^d \delta^d(k - k_1 - k_2)}{\left(m_R^2 - k^2 + i0^+\right) \left(k_1^2 - m_R^2 + i0^+\right)^2 \left(k_2^2 - m_R^2 + i0^+\right)}, \\
\psi_{\Sigma (2)}(\vec{k}_1, \vec{k}_2) &= \psi_{\Sigma (1)}(\vec{k}_2, \vec{k}_1).
\end{aligned}
\label{eq:psi-Sigma1+2}
\end{equation}

Our method, directly inspired by Ref.~\cite{Kovchegov:2012mbw}, proceeds as follows. We begin by replacing $\Gamma_{3R}$ and $\Sigma_R$ by their fully expanded expressions given by the covariant Feynman rules. We assign distinct labels to the momenta of all propagators and keep explicit the integrals over these momenta, along with the energy-momentum conservation factors at each vertex. We then employ an integral representation for the energy conservation factors, expressed~as
\begin{equation}
2\pi\,\delta(k_\text{in}^--k_\text{out}^-) = \int dx^+ e^{-ix^+(k_\text{in}^--k_\text{out}^-)},
\label{eq:delta-energy-conservation}
\end{equation}
where \(k_\text{in}^-\) and \(k_\text{out}^-\) denote the sums of the incoming and outgoing light-cone energies flowing from in to out at the vertex under consideration, respectively. The integration variable $x^+$ is interpreted as the light-cone time at which the considered transition occurs.

\subsubsection{Vertex correction}

We first focus on \(\psi_\Gamma\), as defined in Eq.~\eqref{eq:psi-Gamma3}. Substituting \(\Gamma_{3R}\) with its expression from Eq.~\eqref{eq:def-Gamma3R}, where \(\Gamma_3\) is obtained from the Feynman diagram evaluation as in Eq.~\eqref{eq:Gamma3-from-diagram} but now with all integrals and vertex energy-momentum conservation factors explicitly included, we obtain:
\begin{multline}
\psi_{\Gamma}(\vec{k}_1, \vec{k}_2) = \int \frac{d^d k}{(2\pi)^d} \phi(\vec{k}) \int_{-\infty}^{+\infty} \frac{dk_1^-}{2\pi} 2k_1^+ \int_{-\infty}^{+\infty} \frac{dk_2^-}{2\pi} 2k_2^+
\frac{-\lambda_R}{\left(m_R^2 - k^2 + i0^+\right) \left(k_1^2 - m_R^2 + i0^+\right) \left(k_2^2 - m_R^2 + i0^+\right)} \\
\times \left[ \left(\left.Z_\lambda\right|_{\text{1 loop}} - 1\right) (2\pi)^d \delta^d(k - k_1 - k_2) \right. \\
+ \left. i\lambda_R^2 \int \frac{d^d l}{(2\pi)^d} \frac{d^d k_1'}{(2\pi)^d} \frac{d^d k_2'}{(2\pi)^d} \frac{(2\pi)^d \delta^d(k - k_1' - k_2') (2\pi)^d \delta^d(k_1' + l - k_1) (2\pi)^d \delta^d(k_2' - l - k_2)}{(l^2 - m_R^2 + i0^+) (k_1'^2 - m_R^2 + i0^+) (k_2'^2 - m_R^2 + i0^+)} \right].
\end{multline}

The term proportional to \((Z_\lambda - 1)\), denoted as \(\psi_\Gamma^\text{ct}\), corresponds to the second diagram in Fig.~\ref{subfig:scalarvertex}. This term is simply the renormalization factor multiplied by the tree-level diagram expression discussed earlier in Eq.~\eqref{eq:scalar-tree}. Thus, the equivalence between the covariant formula and the result from light-cone perturbation theory is already established for this contribution.

We now turn to the remaining term, \(\psi_\Gamma^\text{bare}\), whose corresponding Feynman diagram is the first one in Fig.~\ref{subfig:scalarvertex}. We start by expressing the momenta in terms of their light-cone components. We then integrate over the \((d-1)\)-momenta \(\vec{k}\), \(\vec{k_1'}\), and \(\vec{k_2'}\) using the Dirac \(\delta\) factors, and express the light-cone energy-conservation \(\delta\)-factors using Eq.~\eqref{eq:delta-energy-conservation}. This leads to
\begin{multline}
\psi_\Gamma^\text{bare}(\vec{k}_1, \vec{k}_2) = \frac{\phi(\vec{k}_1 + \vec{k}_2)}{2(k_1^+ + k_2^+)} (-i\lambda_R^3) \int \frac{d^{d-2} l^\perp}{(2\pi)^{d-1}} \int_{-\infty}^{+\infty} \frac{dl^+}{8 l^+ (k_1^+ - l^+) (k_2^+ + l^+)}
\int dx_0^+ dx_1^+ dx_2^+\\
\times\int \frac{dk^-}{2\pi} \frac{dk_1^-}{2\pi} \frac{dk_2^-}{2\pi} \frac{dl^-}{2\pi} \frac{dk_1'^-}{2\pi} \frac{dk_2'^-}{2\pi}
\frac{e^{-ix_0^+ (k^- - k_1'^- - k_2'^-)} e^{-ix_1^+ (k_1'^- + l^- - k_1^-)} e^{-ix_2^+ (k_2'^- - l^- - k_2^-)}}{\left(E_R(\vec{k}_1 + \vec{k}_2) - k^- + i0^+\right) \left(k_1^- - \tilde{k}_1^- + i0^+\right) \left(k_2^- - \tilde{k}_2^- + i0^+\right)} \\
\times \frac{1}{\left(l^- - \tilde{l}^- + \sgn(l^+)i{0^+}\right) \left(k_1'^- - E_R(\vec{k}_1 - \vec{l}) + \sgn(k_1^+ - l^+)i{0^+}\right) \left(k_2'^- - E_R(\vec{k}_2 + \vec{l}) + \sgn(k_2^+ + l^+)i{0^+}\right)}.
\label{eq:vertex-bare-1-loop}
\end{multline}
At this point, we must note that spurious, unregularized singularities have been introduced in the integral over \(l^+\), manifesting as poles at \(l^+ = k_1^+\) and \(l^+ = -k_2^+\). Such singularities are typical in LCPT, though absent in the covariant formalism. They generally require {\it ad hoc} regularization when they appear in the LCPT series. (Such regularization is actually not required for the particular diagrams considered in this work.)

The next step involves integrating over the ``\(-\)'' components of the momenta using Cauchy's theorem on a closed contour enclosing the appropriate complex half-plane. We first observe that the integrals over \(k^-\), \(k_1^-\), and \(k_2^-\) are non-zero only if \(x_0^+\), \(x_1^+\), and \(x_2^+\) are strictly negative, ensuring convergence of these integrals at infinity in the upper-half complex plane for \(k^-\), and in the lower-half complex planes for \(k_1^-\) and \(k_2^-\). Hence the integration regions in the three $x^+$ variables are now restricted to $(-\infty,0]$.

For the remaining energy integrals over \(l^-\), \(k_1'^-\), and \(k_2'^-\), we distinguish the following cases:
\begin{enumerate}[(a)]
    \item \(l^+ > 0\): The poles in \(l^-\) and \(k_2'^-\) are located in the lower-half complex plane. To ensure convergence of the phase factor along a contour deformed to enclose these poles, the ordering \(x_1^+ > x_2^+ > x_0^+\) must be satisfied. As a result, \(k_1^+\) must exceed \(l^+\), ensuring that the pole in \(k_1'^-\) also lies in the lower-half complex plane and thus yields a non-zero integral over this variable.

    \item \(l^+ < 0\): The pole in \(l^-\) is located in the upper-half plane, while the pole in \(k_1'^-\) is in the lower-half plane. The only valid ordering is \(x_2^+ > x_1^+ > x_0^+\). Consequently, \(k_2^+\) must exceed \(-l^+\) to ensure that the pole in \(k_2'^-\) lies in the lower-half complex plane, thereby yielding a non-zero integral over this variable.
\end{enumerate}

We split the integration domain of \(l^+\) which contributes to the integral into the union of the intervals \([0, k_1^+]\) and \([-k_2^+, 0]\), effectively decomposing \(\psi_\Gamma^\text{bare}\) into a sum \(\psi_{\Gamma\text{(a)}}^\text{bare} + \psi_{\Gamma\text{(b)}}^\text{bare}\). We then set the appropriate integration boundaries on \(x_0^+\), \(x_1^+\), and \(x_2^+\) in each term and perform the change of variables
\be
(x_0^+,x_1^+,x_2^+)\to
\begin{cases}
(x_1^+,x_{21}^+,x_{02}^+)& \text{in case (a)},\\
(x_2^+,x_{12}^+,x_{01}^+)& \text{in case (b)},
\end{cases}
\label{eq:change-of-var-x+}
\ee
where $x_{ij}^+\equiv x_i^+-x_j^+$, so that the integration domains in the new variables all coincide with $(-\infty,0]$. The terms that depend on the $x^+$ variables in Eq.~(\ref{eq:vertex-bare-1-loop}) are phase factors, that can be grouped as $e^{-i\theta}$, with {\it a priori}
\be
\theta\equiv x_0^+ (k^- - k_1'^- - k_2'^-)+x_1^+ (k_1'^- + l^- - k_1^-)+x_2^+ (k_2'^- - l^- - k_2^-).
\ee
The latter must be rewritten in terms of the variables defined in Eq.~\eqref{eq:change-of-var-x+}. To this end, we introduce distinct notations for the phase \(\theta\) following the two relevant changes of variables:
\be
\begin{aligned}
\theta_\text{(a)}&\equiv x_{02}^+ (k^- - k_1'^- - k_2'^-)+x_{21}^+ (k^--k_1'^- - l^- - k_2^-)+x_1^+ (k^--k_1^--k_2^-),\\
\theta_\text{(b)}&\equiv x_{01}^+ (k^- - k_1'^- - k_2'^-)+x_{12}^+ (k^--k_2'^- + l^- - k_1^-)+x_2^+ (k^--k_1^--k_2^-).
\end{aligned}
\ee

We can now perform the integrals over all ``$-$'' components using the Cauchy theorem. We are left with the residue at the poles
\begin{equation}
\begin{aligned}
& k^- = E_R(\vec{k}_1 + \vec{k}_2) + i0^+, \quad k_1^- = \tilde{k}_1^- - i0^+, \quad k_2^- = \tilde{k}_2^- - i0^+, \\
& k_1'^- = E_R(\vec{k}_1 - \vec{l}) - i0^+, \quad k_2'^- = E_R(\vec{k}_2 + \vec{l}) - i0^+, \quad l^- = \begin{cases}
\tilde{l}^- - i0^+ & \text{in case (a)}, \\
\tilde{l}^- + i0^+ & \text{in case (b)}.
\end{cases}
\end{aligned}
\end{equation}
The integrals then read
\begin{equation}
\begin{aligned}
\psi_{\Gamma\text{(a)}}^\text{bare}(\vec k_1,\vec k_2) &= \frac{\phi(\vec{k}_1 + \vec{k}_2)}{2(k_1^+ + k_2^+)} {i\lambda_R^3} \int \frac{\widetilde{dl}\,{\mathbbm 1}_{\{k_1^+>l^+\}}}{4(k_1^+ - l^+)(k_2^+ + l^+)} \int_{-\infty}^0 dx_1^+  dx_{02}^+  dx_{21}^+ \, e^{-i\theta_{\text{(a)}}(\vec k_1,\vec k_2,\vec l;x_1^+,x_{02}^+,x_{21}^+)}, \\
\psi_{\Gamma\text{(b)}}^\text{bare}(\vec k_1,\vec k_2) &= \frac{\phi(\vec{k}_1 + \vec{k}_2)}{2(k_1^+ + k_2^+)} {i\lambda_R^3} \int \frac{\widetilde{dl}\,{\mathbbm 1}_{\{k_2^+<l^+\}}}{4(k_1^+ + l^+)(k_2^+ - l^+)} \int_{-\infty}^0 dx_2^+  dx_{12}^+  dx_{01}^+ \, e^{-i\theta_{\text{(b)}}(\vec k_1,\vec k_2,\vec l;x_2^+,x_{12}^+,x_{01}^+)},
\end{aligned}
\end{equation}
where the phases now consist of combinations of energy denominators:
\begin{equation}
\begin{aligned}
\theta_{\text{(a)}} &\equiv x_{02}^+ \left[\Delta_R(\vec{k}_1 - \vec{l}, \vec{k}_2 + \vec{l})+i0^+\right] + x_{21}^+ \left[\Delta_R(\vec{k}_1 - \vec{l}, \vec{l}, \vec{k}_2)+i0^+\right] + x_1^+ \left[\Delta_R(\vec{k}_1, \vec{k}_2)+i0^+\right], \\
\theta_{\text{(b)}} &\equiv x_{12}^+ \left[\Delta_R(\vec{k}_1 + \vec{l}, \vec{k}_2 - \vec{l})+i0^+\right] + x_{12}^+ \left[\Delta_R(\vec{k}_1, \vec{l}, \vec{k}_2 - \vec{l})+i0^+\right] + x_2^+ \left[\Delta_R(\vec{k}_1, \vec{k}_2)+i0^+\right].
\end{aligned}
\end{equation}
We have performed the change of variable \(\vec{l} \to -\vec{l}\) in the integrand of \(\psi_{\Gamma\text{(b)}}^\text{bare}\), and consequently in the phase \(\theta_{\text{(b)}}\).

Finally, integrating over the "+" components generates three energy denominators:
\begin{equation}
\begin{aligned}
\psi_{\Gamma\text{(a)}}^\text{bare}(\vec k_1,\vec k_2) &= \frac{\phi(\vec{k}_1 + \vec{k}_2)}{2(k_1^+ + k_2^+)} \frac{\lambda_R^3}{\Delta_R(\vec{k}_1, \vec{k}_2)} \int \frac{\widetilde{dl}\,{\mathbbm 1}_{\{k_1^+>l^+\}}}{4(k_1^+ - l^+)(k_2^+ + l^+)} \frac{1}{\Delta_R(\vec{k}_1 - \vec{l}, \vec{k}_2 + \vec{l}) \Delta_R(\vec{k}_1 - \vec{l}, \vec{l}, \vec{k}_2)}, \\
\psi_{\Gamma\text{(b)}}^\text{bare}(\vec k_1,\vec k_2) &= \frac{\phi(\vec{k}_1 + \vec{k}_2)}{2(k_1^+ + k_2^+)} \frac{\lambda_R^3}{\Delta_R(\vec{k}_1, \vec{k}_2)} \int \frac{\widetilde{dl}\,{\mathbbm 1}_{\{k_2^+<l^+\}}}{4(k_1^+ + l^+)(k_2^+ - l^+)} \frac{1}{\Delta_R(\vec{k}_1 + \vec{l}, \vec{k}_2 - \vec{l}) \Delta_R(\vec{k}_1, \vec{l}, \vec{k}_2 - \vec{l})}.
\end{aligned}
\end{equation}

\begin{figure}
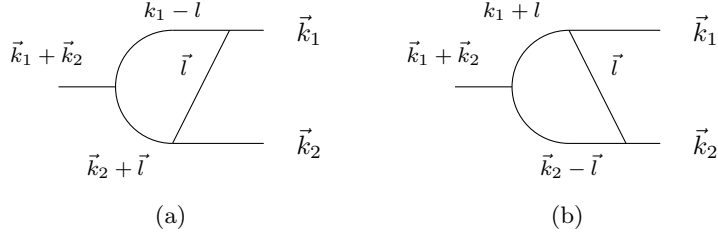

    \begin{center}
    \begin{subfigure}[b]{.3\textwidth}
        \centerline{\LCtrianglebis}
        \caption{}
        \label{subfig:vertex-a}
    \end{subfigure}
    \begin{subfigure}[b]{.3\textwidth}
        \centerline{\LCtriangle}
        \caption{}
        \label{subfig:vertex-b}
    \end{subfigure}
    \end{center}
    \caption{One-loop vertex correction diagrams in light-cone perturbation theory, that correspond to the covariant diagram of Fig.~\ref{subfig:scalarvertex}. Light-cone time increases from left to right, and the $+$ components of the displayed momenta are all positive.}
    \label{fig:triangle-graphs}
\end{figure}

The terms (a) and (b) correspond to the contributions of the time-ordered diagrams in Figs.~\ref{subfig:vertex-a} and~\ref{subfig:vertex-b}, respectively, evaluated using the rules of light-cone perturbation theory. For example, the expression for \(\psi_{\Gamma\text{(a)}}^\text{bare}\) originates from the term
\begin{equation}
\int \widetilde{dk} \, \frac{\sqrt{Z_\varphi} \, \phi(\vec{k})}{\tilde{k}^- - \tilde{k}_1^- - \tilde{k}_2^-} \int \widetilde{dk_1'} \widetilde{dk_2'} \widetilde{dl} \frac{\mel{\vec{k}_1}{\mathcal{H}_{1R}^\text{bare}}{\vec{k}_1', \vec{l}} \mel{\vec{k}_2, \vec{l}}{\mathcal{H}_{1R}^\text{bare}}{\vec{k}_2'} \mel{\vec{k}_1', \vec{k}_2'}{\mathcal{H}_{1R}^\text{bare}}{\vec{k}}}{\left(\tilde{k}^- - \tilde{k}_1'^- - \tilde{k}_2^- - \tilde{l}^-\right) \left(\tilde{k}^- - \tilde{k}_1'^- - \tilde{k}_2'^-\right)},
\end{equation}
in the LCPT series~\eqref{eq:LCPT}, with \(\sqrt{Z_\varphi} \to 1\), \(j = 2\), \(\ket{\Phi_1} = \ket{\vec{k}_1', \vec{k}_2'}\), and \(\ket{\Phi_2} = \ket{\vec{k}_1', \vec{l}, \vec{k}_2}\).\footnote{The combinatorial factors arising from the reduction of the matrix elements of \(\mathcal{H}_{1R}^\text{bare}\) to the elementary \(1 \to 2\) and \(2 \to 1\) transition amplitudes cancel the factorials in the denominators of Eq.~\eqref{eq:LCPT}.} This matches Eq.~(140) in Ref.~\cite{Munier:2025qyz}.

\subsubsection{Self-energy diagrams}

The contribution \(\psi_{\Sigma (0)}\), as defined in Eq.~\eqref{eq:psi-Sigma0}, consists of the normalization factor \(\left(\sqrt{Z_\varphi} - 1\right)\) evaluated at one-loop accuracy, multiplied by the tree-level expression given in Eq.~\eqref{eq:scalar-tree}. Consequently, it is evident that our formula yields the same result as the application of the LCPT rules.

We just need to discuss \(\psi_{\Sigma (1)}\), as defined in Eq.~\eqref{eq:psi-Sigma1+2}. Substituting the self-energy with the one-loop expression in Eq.~\eqref{eq:Sigma-from-diagram} to which one adds the contribution of the counter-term graph, we obtain:
\begin{multline}
\psi_{\Sigma (1)}(\vec k_1,\vec k_2) = \int \frac{d^d k}{(2\pi)^d} \phi(\vec{k}) \int_{-\infty}^{+\infty} \frac{dk_1^-}{2\pi} 2k_1^+ \int_{-\infty}^{+\infty} \frac{dk_2^-}{2\pi} 2k_2^+ \frac{-\lambda_R}{\left(m_R^2 - k^2 + i0^+\right) \left(k_1^2 - m_R^2 + i0^+\right) \left(k_2^2 - m_R^2 + i0^+\right)} \\
\times \int \frac{d^d k_1'}{(2\pi)^d} \frac{(2\pi)^d \delta^d(k - k_1' - k_2)}{k_1'^2 - m_R^2 + i0^+} \bigg(
\frac{i\lambda_R^2}{2} \int \frac{d^d l_1}{(2\pi)^d} \frac{d^d l_2}{(2\pi)^d} \frac{(2\pi)^d \delta^d(k_1' - l_1 - l_2) (2\pi)^d \delta^d(l_1 + l_2 - k_1)}{(l_1^2 - m_R^2 + i0^+) (l_2^2 - m_R^2 + i0^+)}\\
+(Z_m - 1) m_R^2 (2\pi)^d \delta^d(k_1' - k_1)
\bigg).
\end{multline}
This expression contains two terms, corresponding to the two diagrams in Fig.~\ref{subfig:scalarselfenergyone}. We denote them as \(\psi_{\Sigma (1)}^\text{bare}\) and \(\psi_{\Sigma (1)}^\text{ct}\), respectively.

\paragraph{Bare diagram}

We now focus on \(\psi_{\Sigma (1)}^\text{bare}\). We follow a procedure similar to that used for the vertex correction term \(\psi_\Gamma^\text{bare}\). Specifically, we perform the integrals over the \((d-1)\)-momenta and express the factors enforcing the conservation of the ``\(-\)'' components using Eq.~\eqref{eq:delta-energy-conservation}. This leads to
\begin{multline}
\psi_{\Sigma (1)}^\text{bare}(\vec k_1,\vec k_2) = \frac{\phi(\vec{k}_1 + \vec{k}_2)}{2(k_1^+ + k_2^+)} \frac{(-i\lambda_R^3)}{4 k_1^+} \int dx_0^+ \, dx_1'^+ \, dx_1^+ \int_{-\infty}^{+\infty} \frac{dk^-}{2\pi} \frac{dk_1^-}{2\pi} \frac{dk_2^-}{2\pi} \frac{dk_1'^-}{2\pi} \\
\times \frac{1}{\left(E_R(\vec{k}_1 + \vec{k}_2) - k^- + i0^+\right) \left(k_1'^- - \tilde{k}_1^- + i0^+\right) \left(k_1^- - \tilde{k}_1^- + i0^+\right) \left(k_2^- - \tilde{k}_2^- + i0^+\right)} \\
\times \int \frac{d^{d-1} \vec l_1}{(2\pi)^{d-1} 4 l_1^+ (k_1^+ - l_1^+)} \int \frac{dl_1^-}{2\pi} \frac{dl_2^-}{2\pi} \frac{e^{-ix_0^+ (k^- - k_1'^- - k_2^-)} e^{-ix_1'^+ (k_1'^- - l_1^- - l_2^-)} e^{-ix_1^+ (l_1^- + l_2^- - k_1^-)}}{\left(l_1^- - \tilde{l}_1^- + \sgn(l_1^+)i{0^+}\right) \left(l_2^- - E_R(\vec{k}_1 - \vec{l}_1) + \sgn(k_1^+ - l_1^+)i{0^+}\right)}.
\end{multline}

Following the same reasoning as for the vertex corrections, we observe that only negative values of \(x_0^+\), \(x_1'^+\), and \(x_1^+\) contribute to the integrals. Additionally, \(l_1^+\) and \(k_1^+ - l_1^+\) must share the same sign for the integrals over \(l_1^-\) and \(l_2^-\) to yield non-zero results. Since these two quantities cannot both be negative, we necessarily have \(k_1^+ > l_1^+ > 0\). This implies the ordering \(x_0^+ < x_1'^+ < x_1^+\) of the light-cone times.

We then rewrite the phase factors in the integrand, grouped as \(e^{-i\theta}\), using more suitable variables: \(x_{01'}^+ \equiv x_0^+ - x_1'^+\), \(x_{1'1}^+ \equiv x_1'^+ - x_1^+\), and \(x_1^+\), all ranging in \((-\infty, 0]\). The phase can be rearranged as
\begin{equation}
\begin{aligned}
\theta &= x_0^+ (k^- - k_1'^- - k_2^-) + x_1'^+ (k_1'^- - l_1^- - l_2^-) + x_1^+ (l_1^- + l_2^- - k_1^-) \\
&= x_{01'}^+ (k^- - k_1'^- - k_2^-) + x_{1'1}^+ (k^- - k_2^- - l_1^- - l_2^-) + x_1^+ (k^- - k_1^- - k_2^-).
\end{aligned}
\end{equation}

We are now ready to perform all integrals over the ``\(-\)'' components by Cauchy's theorem. It leaves us with the residue at the poles located at
\begin{equation}
\begin{aligned}
k^- &= E_R(\vec{k}_1 + \vec{k}_2) + i0^+, \quad k_1^- = k_1'^- = \tilde{k}_1^- - i0^+, \\
k_2^- &= \tilde{k}_2^- - i0^+, \quad l_1^- = \tilde{l}_1^- - i0^+, \quad l_2^- = E_R(\vec{k}_1 - \vec{l}_1) - i0^+.
\end{aligned}
\end{equation}
Putting back all factors, we obtain:
\begin{multline}
\psi_{\Sigma (1)}^\text{bare}(\vec k_1,\vec k_2) = \frac{\phi(\vec{k}_1 + \vec{k}_2)}{2(k_1^+ + k_2^+)} \frac{i\lambda_R^3}{2} \int \frac{\widetilde{dl_1}\,{\mathbbm 1}_{\{k_1^+>l_1^+\}}}{4k_1^+(k_1^+ - l_1^+)} \int_{-\infty}^{0} dx_{01'}^+ \, dx_{1'1}^+ \, dx_1^+ \\
\times e^{-ix_{01'}^+ \left(\Delta_R(\vec{k}_1, \vec{k}_2) + i0^+\right)} e^{-ix_{1'1}^+ \left(\Delta_R(\vec{l}_1, \vec{k}_1 - \vec{l}_1, \vec{k}_2) + i0^+\right)} e^{-ix_1^+ \left(\Delta_R(\vec{k}_1, \vec{k}_2) + i0^+\right)}.
\end{multline}
Finally, integrating over the light-cone times, we find:
\begin{equation}
\psi_{\Sigma (1)}^\text{bare}(\vec k_1,\vec k_2) = \frac{\phi(\vec{k}_1 + \vec{k}_2)}{2(k_1^+ + k_2^+)} \frac{\lambda_R^3}{2 \left[\Delta_R(\vec{k}_1, \vec{k}_2)\right]^2} \int \frac{\widetilde{dl_1}\,{\mathbbm 1}_{\{k_1^+>l_1^+\}}}{4k_1^+(k_1^+ - l_1^+)} \frac{1}{\Delta_R(\vec{l}_1, \vec{k}_1 - \vec{l}_1, \vec{k}_2)}.
\end{equation}
This expression is identical to that found using LCPT; compare to Eq.~(127) in Ref.~\cite{Munier:2025qyz}.


\paragraph{Counter-term diagram}

For \(\psi_{\Sigma (1)}^\text{ct}\), we perform the integrations over \(k\) and \(k_1'\) using the Dirac energy-momentum conservation factors. This leaves us with
\begin{equation}
\psi_{\Sigma (1)}^\text{ct}(\vec k_1,\vec k_2) = \frac{\phi(\vec{k}_1 + \vec{k}_2)}{2(k_1^+ + k_2^+)} \int_{-\infty}^{\infty} \frac{dk_1^-}{2\pi} \frac{dk_2^-}{2\pi} \frac{(Z_m - 1) m_R^2}{\left(E_R(\vec{k}_1 + \vec{k}_2) - k_1^- - k_2^- + i0^+\right) (k_1^- - \tilde{k}_1^- + i0^+)^2 (k_2^- - \tilde{k}_2^- + i0^+)}.
\end{equation}

We evaluate the remaining integrals using Cauchy's theorem. We start with the integral over \(k_1^-\), selecting the unique simple pole in the upper-half complex plane at \(k_1^- = E_R(\vec{k}_1, \vec{k}_2) - k_2^- + i0^+\). This leaves us with the integral over \(k_2^-\), which has a unique simple pole in the lower-half complex plane at \(k_2^- = \tilde{k}_2^- - i0^+\). We finally obtain:
\begin{equation}
\psi_{\Sigma (1)}^\text{ct}(\vec k_1,\vec k_2) = \frac{\phi(\vec{k}_1 + \vec{k}_2)}{2(k_1^+ + k_2^+)} \frac{(Z_m - 1) m_R^2}{\left(\Delta_R(\vec{k}_1, \vec{k}_2)\right)^2}.
\end{equation}
This result is exactly what one would obtain using the rules of LCPT applied to the second diagram in Fig.~\ref{subfig:scalarselfenergyone}, viewed as a time-ordered diagram.

\subsubsection{\texorpdfstring{Insights from the verification of the formula~\eqref{eq:relation-scalar}}{Insights into the formula}}

Let us draw a few conclusions from our explicit calculations regarding Eq.~\eqref{eq:relation-scalar}:
\begin{itemize}
    \item The sign change of the small imaginary regulator \(i0^+\) in the propagator of the incoming particle, with respect to the standard Feynman regulator, was crucial. It ensured that all interactions occurred at negative light-cone times, thereby correctly generating the appropriate energy denominators.
    \item The use of a bare propagator for the initial particle is justified by the absence of diagrams representing self-energy contributions to the one-particle initial state in the series~\eqref{eq:LCPT}. These contributions are effectively accounted for by the wave function renormalization factor \(\sqrt{Z_\varphi}\). On the light-cone side, a different formulation of LCPT would be required to obtain a diagrammatic representation of this factor, as discussed in Ref.~\cite{Munier:2025qyz}. However, the corresponding covariant formula remains unknown to us.
\end{itemize}

We expect this approach to generalize to higher orders, since the mechanism described above, namely closing contours to pick up residues and discontinuities that reconstruct the LCPT energy denominators, does not rely on any feature specific to single loop diagrams. However, we have not established this in full generality.


\subsection{Covariant evaluation of the wave function}
\label{sec:re-computation}

Having established the equivalence of Eq.~(\ref{eq:relation-scalar}) with light-cone perturbation theory calculation of the wave functions up to one-loop accuracy, we now evaluate the individual terms in Eq.~\eqref{eq:relation-scalar-1-loop} using the covariant expressions for \(\Gamma_{3R}\) and \(\Sigma_R\). For this purpose, it is convenient to begin with the one-loop expansion of Eq.~(\ref{eq:relation-scalar-integrated-integrated}). Let us enumerate the distinct contributions in Eq.~(\ref{eq:relation-scalar-1-loop}). Beginning with the vertex correction term,
\be
\psi_{\Gamma}(\vec k_1,\vec k_2)=\frac{\phi(\vec k_1+\vec k_2)}{2(k_1^++k_2^+)}\int\frac{dk_1^-}{2i\pi}
\frac{\left.\Gamma_{3R}\left[m_R^2,2k_1^+(k_1^--\tilde k_1^-)+m_R^2,2k_2^+\left(\Delta_R(\vec k_1,\vec k_2)+\tilde k_1^--k_1^-\right)+m_R^2\right]\right|_{{\cal O}(\bar\lambda_R^3)}}{\left(k_1^--\tilde k_1^-+i0^+\right)\left(\Delta_R(\vec k_1,\vec k_2)+\tilde k_1^--k_1^-+i0^+\right)}\,.
\label{eq:vertex-1-loop-integrated}
\ee
As for the self-energy contributions,
\be
\begin{aligned}
\psi_{\Sigma(0)}(\vec k_1,\vec k_2)&=\frac{\phi(\vec k_1+\vec k_2)}{2\,(k_1^++k_2^+)}
\int_{-\infty}^{\infty}\frac{dk_1^-}{2i\pi}\,
\frac{-\lambda_R\left(\sqrt{Z_\varphi}\big|_{\text{1 loop}}-1\right)}{\left(k_1^--\tilde k_1^-+i0^+\right)\left(\Delta_R(\vec k_1,\vec k_2)+\tilde k_1^--k_1^-+i0^+\right)}\,,\\
\psi_{\Sigma(1)}(\vec k_1,\vec k_2)&=\frac{\phi(\vec k_1+\vec k_2)}{2\,(k_1^++k_2^+)}
\int_{-\infty}^{\infty}\frac{dk_1^-}{2i\pi}\,
\frac{-\lambda_R\Sigma_R\left(2k_1^+(k_1^--\tilde k_1^-)+m_R^2\right)\big|_{\text{1 loop}}}{2k_1^+\left(k_1^--\tilde k_1^-+i0^+\right)^2\left(\Delta_R(\vec k_1,\vec k_2)+\tilde k_1^--k_1^-+i0^+\right)}\,,\\
\psi_{\Sigma(2)}(\vec k_1,\vec k_2)&=\frac{\phi(\vec k_1+\vec k_2)}{2\,(k_1^++k_2^+)}
\int_{-\infty}^{\infty}\frac{dk_1^-}{2i\pi}\,
\frac{-\lambda_R\Sigma_R\left[2k_2^+\left(\Delta_R(\vec k_1,\vec k_2)+\tilde k_1^--k_1^-\right)+m_R^2\right]\big|_{\text{1 loop}}}{2k_2^+\left(k_1^--\tilde k_1^-+i0^+\right)\left(\Delta_R(\vec k_1,\vec k_2)+\tilde k_1^--k_1^-+i0^+\right)^2}\,.
\end{aligned}
\label{eq:self-energy-1-loop-integrated}
\ee


\subsubsection{Vertex correction diagram}

\begin{figure}
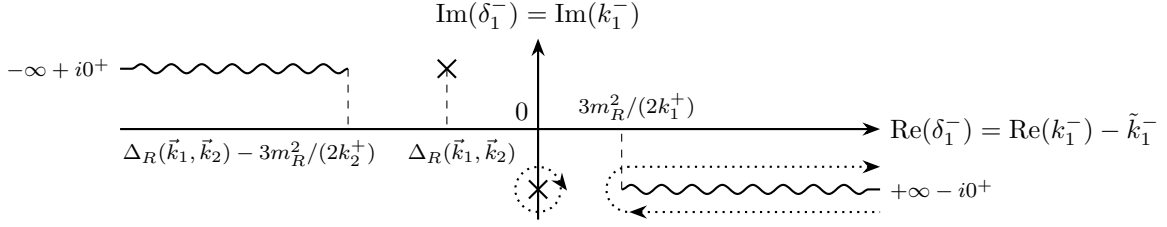

    \centering
    \contourVertex
    \caption{Complex plane of the \(k_1^-\) variable for the diagram in Fig.~\ref{subfig:scalarvertex}. The initial integration contour runs along the entire real axis. We turn this contour into a closed contour enclosing the lower half-plane, which can then be split and shrunk to surround a single pole and a cut (dotted lines). The integral~\eqref{eq:vertex-1-loop-integrated} then reduces to the sum of the residue at this pole and the integral of the discontinuity of the integrand across the cut.}
    \label{fig:complex-plane-k1--vertex}
\end{figure}

We first focus on the vertex correction term given in Eq.~(\ref{eq:vertex-1-loop-integrated}). 

We need to perform the integral over $k_1^-$. The analytic structure of the integrand is shown in Fig.~\ref{fig:complex-plane-k1--vertex}. Let us close the contour in the lower half-plane. There is a contribution from the simple pole at $k_1^-=\tilde k_1^--i0^+$, another one from the cut $k_1^-\in[3m_R^2/(2k_1^+),+\infty)$:
\begin{multline}
\psi_{\Gamma}(\vec k_1,\vec k_2)=\frac{\phi(\vec k_1+\vec k_2)}{2(k_1^++k_2^+)}
\Bigg(
-\frac{\left.\Gamma_{3R}\left(m_R^2,m_R^2,2k_2^+\Delta_R(\vec k_1,\vec k_2)+m_R^2\right)\right|_{{\cal O}(\bar\lambda_R^3)}}{\Delta_R(\vec k_1,\vec k_2)}\\
+\int_{\frac{3m_R^2}{2k_1^+}}^{+\infty}\frac{d\delta_1^-}{2i\pi}
\frac{\left.\text{Disc}_{\delta_1^-}\Gamma_{3R}\left[m_R^2,2k_1^+\delta_1^-+m_R^2,2k_2^+\left(\Delta_R(\vec k_1,\vec k_2)-\delta_1^-\right)+m_R^2\right]\right|_{{\cal O}(\bar\lambda_R^3)}}{\delta_1^-\left(\delta_1^--\Delta_R(\vec k_1,\vec k_2)\right)}\Bigg),
\label{eq:psi-Gamma3R-massive}
\end{multline}
where we performed the change of variable $k_1^-\to \delta_1^-\equiv k_1^--\tilde k_1^-$ in the integral in the second term.

At this point, it is in order to discuss the sign of the arguments of \(\Gamma_{3R}\). Simple kinematic considerations reveal that
\be
2k_i^+\Delta_R(\vec{k}_1, \vec{k}_2) + m_R^2 \leq 0
\quad \text{for } i = 1, 2.
\label{eq:sign-Delta}
\ee
Indeed, the left-hand side of this inequality corresponds precisely to the virtuality of particle \(i=1,2\) when the other two particles, the initial one and the other final one, are placed on their mass shell. In such a configuration, the virtuality of particle \(i\) is necessarily negative. Hence the $\Gamma_{3R}$'s that appear in Eq.~(\ref{eq:psi-Gamma3R-massive}) all have their first two parameters positive, and the last one negative.

We could now substitute the known expressions for \(\Gamma_{3R}\) and its relevant discontinuity into this formula. Up to a single well-behaved one-dimensional integration, this would directly yield a particularly compact expression for the corresponding contribution to the wave function. Instead, let us verify this expression in the massless limit $m_R\to 0$, which is meaningful for this particular contribution to the wave function and simplifies the calculations significantly.


\paragraph{Massless limit}

We set $m_R=0$ in Eq.~\eqref{eq:psi-Gamma3R-massive}. Observing that $k_2^2\equiv 2k_2^+\Delta_R(\vec k_1,\vec k_2)$ is always negative (see Eq.~\eqref{eq:sign-Delta}), we can use the calculations performed in Appendix~\ref{sec:appendix-triangle}. The first term in Eq.~\eqref{eq:psi-Gamma3R-massive} is given by Eq.~\eqref{eq:Gamma3R-real} with $k_1^2=0$ and $k_2^2$ set to $2k_2^+\Delta_R(\vec k_1,\vec k_2)$. As for the second term, the discontinuity in the integrand is deduced from Eq.~\eqref{eq:Gamma3R-Im} up to the replacements $k_1^2\to 2k_1^+\delta_1^-$ and $k_2^2\to 2k_2^+\left(\Delta_R(\vec k_1,\vec k_2)-\delta_1^-\right)$. Hence
\begin{multline}
    \psi_{\Gamma}(\vec k_1,\vec k_2)=-\frac{\phi(\vec k_1+\vec k_2)}{2(k_1^++k_2^+)}
\frac{\bar\lambda_R^3}{2(4\pi)^3}\Bigg[
\frac{1}{\Delta_R(\vec k_1,\vec k_2)}\left(\ln\frac{-2k_2^+\Delta_R(\vec k_1,\vec k_2)}{\mu^2}-3\right)\\
+\int_{0}^{+\infty}{d\delta_1^-}
\frac{k_1^+}{\left[\left(k_1^++k_2^+\right)\delta_1^--k_2^+\Delta_R(\vec k_1,\vec k_2)\right]\left[\delta_1^--\Delta_R(\vec k_1,\vec k_2)\right]}\Bigg].
\label{eq:psi-Gamma3R-massless}
\end{multline}
The integral is straightforward to perform. All in all, we get
\be
\psi_{\Gamma}(\vec k_1,\vec k_2)=-\frac{\phi(\vec k_1+\vec k_2)}{2(k_1^++k_2^+)}
\frac{\bar\lambda_R^3}{2(4\pi)^3}\frac{1}{\Delta_R(\vec k_1,\vec k_2)}\left(
\ln\frac{-2(k_1^++k_2^+)\Delta_R(\vec k_1,\vec k_2)}{\mu^2}-3
\right).
\ee
This result matches exactly that obtained from a direct LCPT calculation; see Eq.~(154) in Ref.~\cite{Munier:2025qyz}. While the latter required a more involved calculation and relies on ``unnatural'' cancellations between the two contributing time-ordered diagrams, the present approach yields the same result in a straightforward manner, assuming analytical expressions for \(\Gamma_{3R}\) are available.

\subsubsection{Self-energy diagrams}

\begin{figure}
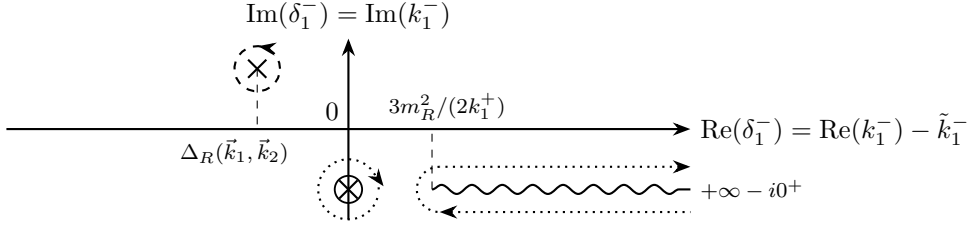

    \centering
    \contourSelf
    \caption{Complex plane of the \(k_1^-\) variable for the diagram in Fig.~\ref{subfig:scalarselfenergyone}. The initial integration contour consists of the entire real axis. It can be deformed in two ways: either to the upper half-plane, enclosing a single pole (dashed line), or to the lower half-plane, enclosing a double pole and a cut (dotted lines).}
    \label{fig:complex-plane-k1--self-1}
\end{figure}

We now analyze the first non-trivial self-energy contribution in Eq.~\eqref{eq:self-energy-1-loop-integrated}, namely $\psi_{\Sigma(1)}(\vec{k}_1, \vec{k}_2)$. The integrand exhibits the analytical structure illustrated in Fig.~\ref{fig:complex-plane-k1--self-1}: there is a single pole in the upper half-plane, corresponding to particle 2 going on-shell, a double pole in the lower half-plane, corresponding to particle 1 going on-shell, and a branch cut in the lower half-plane starting at the two-particle production threshold: \(k_1^2 > 4m_R^2\).

By closing the contour in the upper-half plane, which selects the single pole, we obtain:
\begin{equation}
\psi_{\Sigma (1)}(\vec{k}_1, \vec{k}_2) = \frac{\phi(\vec{k}_1 + \vec{k}_2)}{2(k_1^+ + k_2^+)} \lambda_R \frac{\left.\Sigma_R\left(2k_1^+ \Delta_R(\vec{k}_1, \vec{k}_2) + m_R^2\right)\right|_{\text{1 loop}}}{2k_1^+ \left(\Delta_R(\vec{k}_1, \vec{k}_2)\right)^2}.
\label{eq:psi-Sigma1-res-1-pole}
\end{equation}
In other words, the contribution of the diagrams in Fig.~\ref{subfig:scalarselfenergyone} can be straightforwardly written down once the covariant expression of \(\Sigma_R\) is known.

Substituting \(\Sigma_R\) with its expression from Eq.~\eqref{eq:SigmaR-one-loop}, adding the contributions of $\psi_{\Sigma (0)}$ and $\psi_{\Sigma (2)}$, we verify that the result matches that obtained starting from a LCPT calculation; compare to Eq.~(138) in Ref.~\cite{Munier:2025qyz}.

It is instructive to consider the formal limit \(\Delta_R \to 0\), which corresponds to energy conservation between the asymptotic one-particle state and the two-particle states onto which one projects to obtain the light-cone wave function. In this limit, we have
\be
\frac{\Sigma_R\left(2k_1^+ \Delta_R(\vec{k}_1, \vec{k}_2) + m_R^2\right)}{2k_1^+} = \Sigma_R'(m_R^2) \Delta_R(\vec{k}_1, \vec{k}_2) + o(\Delta_R).
\ee
At one-loop accuracy, \(\Sigma_R'(m_R^2) \simeq Z_\varphi - 1\); see Eq.~\eqref{eq:dressed-prop-near-massshell}. Summing all self-energy contributions together with the tree-level contribution, Eq.~(\ref{eq:scalar-tree}), we may write the result as
\be
\left.\psi_{\phi\to\varphi\varphi}(\vec{k}_1, \vec{k}_2)\right|_\text{tree} + \psi_{\Sigma (0)}(\vec{k}_1, \vec{k}_2) + \psi_{\Sigma (1)}(\vec{k}_1, \vec{k}_2) + \psi_{\Sigma (2)}(\vec{k}_1, \vec{k}_2) \underset{\Delta_R \to 0}{\longrightarrow} \frac{\phi(\vec{k}_1 + \vec{k}_2)}{2(k_1^+ + k_2^+)} \frac{\lambda_R \left(\left.Z_\varphi\right|_{\text{1 loop}}\right)^{5/2}}{\Delta_R(\vec{k}_1, \vec{k}_2)},
\label{eq:self-energies-onshell-limit}
\ee
which is consistent up to one-loop accuracy.
This is the expression~\eqref{eq:scalar-tree} of the tree-level diagram multiplied by one wave-function renormalization factor $\sqrt{Z_\varphi}$ that dresses the incoming particle, and by two more factors $Z_\varphi$ that correspond to the residues at the physical poles of the full propagators of the outgoing particles.

\paragraph{Alternative calculation}

Instead of picking the pole associated with particle 2 going on-shell in the upper half-plane in Fig.~\ref{fig:complex-plane-k1--self-1}, the contour can be closed in the lower half-plane. In this case, it encloses both a double pole and a cut. Specifically, we have:
\begin{equation}
\psi_{\Sigma (1)}(\vec{k}_1, \vec{k}_2) = \frac{\phi(\vec{k}_1 + \vec{k}_2)}{2(k_1^+ + k_2^+)} \lambda_R \left( \frac{\left.\Sigma_R'(m_R^2)\right|_{\text{1 loop}}}{\Delta_R(\vec{k}_1, \vec{k}_2)} + \int_{\frac{3m_R^2}{2k_1^+}}^{+\infty} \frac{d\delta_1^-}{2i\pi} \frac{\left.\text{Disc}_{k_1^-}\Sigma_R\left(2k_1^+ \delta_1^- + m_R^2\right)\right|_{\text{1 loop}}}{2k_1^+ (\delta_1^-)^2 \left(\delta_1^- - \Delta_R(\vec{k}_1, \vec{k}_2)\right)} \right).
\label{eq:psi-Sigma1-alternative}
\end{equation}
The first term corresponds to the contribution from the double pole, while the second term arises from the branch cut.

This expression can be directly matched with Eq.~\eqref{eq:psi-Sigma1-res-1-pole}, as it represents a spectral decomposition of the latter. To explicitly demonstrate this match, we use the dispersion relation provided in Eq.~\eqref{eq:dispersion-Sigma}. Specifically, it suffices to express the invariants in that formula as follows:
\be
\begin{aligned}
k^2 &\to 2k_1^+\Delta_R(\vec{k}_1, \vec{k}_2) + m_R^2\,, \\
s &\to 2k_1^+\delta_1^- + m_R^2\,.
\end{aligned}
\ee
By doing so, we observe that the right-hand side of Eq.~\eqref{eq:dispersion-Sigma} matches the terms inside the parentheses of Eq.~\eqref{eq:psi-Sigma1-alternative}, up to a factor of \(2k_1^+.\) Similarly, the left-hand side matches the corresponding factor in Eq.~\eqref{eq:psi-Sigma1-res-1-pole}, again up to the same factor \(2k_1^+\).


\section{Higher orders}
\label{sec:two-loops}

In the previous section, we have focused on one-loop contributions. However, we can derive expressions for higher-order contributions in the same manner.

\begin{figure}
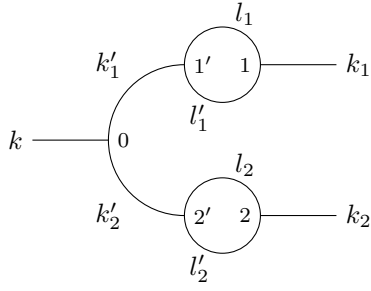

\centerline\scalarselfenergytwo
\caption{A two-loop covariant diagram. The \(d\)-momenta labeling the propagators flow from left to right.}
\label{fig:two-loop}
\end{figure}

Let us apply Eq.~\eqref{eq:relation-scalar-integrated-integrated} to the more complex diagram shown in Fig.~\ref{fig:two-loop}. This diagram features one-loop self-energy insertions on the propagators of both outgoing particles. The combined contribution to the light-cone wave function \(\psi_{\phi\to\varphi\varphi}(\vec{k}_1, \vec{k}_2)\), including the corresponding counter-terms (three diagrams, not shown), is given~by
\begin{multline}
\left.\psi_{\phi\to\varphi\varphi}(\vec{k}_1, \vec{k}_2)\right|_{\text{Fig.~\ref{fig:two-loop}}+\text{ct}} = \frac{\phi(\vec{k}_1 + \vec{k}_2)}{2(k_1^+ + k_2^+)} \\
\times (-\lambda_R) \int \frac{d\delta_1^-}{2i\pi} \frac{\left.\Sigma_R\left(2k_1^+\delta_1^- + m_R^2\right)\right|_{\text{1 loop}}}{2k_1^+ \left(\delta_1^- + i0^+\right)^2}
\frac{\left.\Sigma_R\left[2k_2^+\left(\Delta_R(\vec{k}_1, \vec{k}_2) - \delta_1^-\right) + m_R^2\right]\right|_{\text{1 loop}}}{2k_2^+ \left(\Delta_R(\vec{k}_1, \vec{k}_2) - \delta_1^- + i0^+\right)^2}.
\label{eq:relation-scalar-2-loop}
\end{multline}
The singularities of the integrand are similar to those shown in Fig.~\ref{fig:complex-plane-k1--vertex}, except that the two poles are now double. 

By closing the contour in the lower half-plane, we pick the double pole at \(\delta_1^- = -i0^+\) and the discontinuity of the first self-energy function \(\Sigma_R\):
\begin{multline}
\left.\psi_{\phi\to\varphi\varphi}(\vec{k}_1, \vec{k}_2)\right|_{\text{Fig.~\ref{fig:two-loop}}+\text{ct}} = \frac{\phi(\vec{k}_1 + \vec{k}_2)}{2(k_1^+ + k_2^+)} \lambda_R \Bigg( \frac{\left.\Sigma_R'(m_R^2)\right|_{\text{1 loop}} \left.\Sigma_R\left(2k_2^+\Delta_R(\vec{k}_1, \vec{k}_2) + m_R^2\right)\right|_{\text{1 loop}}}{2k_2^+ \left(\Delta_R(\vec{k}_1, \vec{k}_2)\right)^2} \\
- \int_{\frac{3m_R^2}{2k_1^+}}^{+\infty} \frac{d\delta_1^-}{2i\pi} \frac{\left.\text{Disc}_{\delta_1^-}\Sigma_R\left(2k_1^+ \delta_1^- + m_R^2\right)\right|_{\text{1 loop}}}{2k_1^+ \left(\delta_1^-\right)^2}
\frac{\left.\Sigma_R\left[2k_2^+\left(\Delta_R(\vec{k}_1, \vec{k}_2) - \delta_1^-\right) + m_R^2\right]\right|_{\text{1 loop}}}{2k_2^+ \left(\Delta_R(\vec{k}_1, \vec{k}_2) - \delta_1^-\right)^2} \Bigg).
\label{eq:relation-scalar-2-loop-contour}
\end{multline}
This expression is quite simple, as all elements are elementary functions given by Eqs.~\eqref{eq:SigmaR-one-loop-rewritten} and~\eqref{eq:Sigma-one-loop-disc}, and only a single integration is required. By contrast, using the rules of LCPT would require considering one diagram per vertex ordering in light-cone time, namely six diagrams (three of which have genuinely distinct expressions). For each such diagram, two non-trivial \((d-1)\)-dimensional integrations would be necessary.

This illustrates very well the tremendous simplification brought by using covariant amplitudes.

\paragraph{Restoring the exchange symmetry}

Equation~\eqref{eq:relation-scalar-2-loop-contour} is not manifestly symmetric under the exchange \(k_1 \leftrightarrow k_2\). To obtain a manifestly symmetric expression, we can again use the dispersion relation~\eqref{eq:dispersion-Sigma}. First, we apply it to the \(\Sigma_R\) factor inside the integral:
\begin{multline}
\frac{\left.\Sigma_R\left[2k_2^+\left(\Delta_R(\vec{k}_1, \vec{k}_2) - \delta_1^-\right) + m_R^2\right]\right|_{\text{1 loop}}}{2k_2^+ \left(\Delta_R(\vec{k}_1, \vec{k}_2) - \delta_1^-\right)^2} = \frac{\left.\Sigma_R'(m_R^2)\right|_{\text{1 loop}}}{\Delta_R(\vec{k}_1, \vec{k}_2) - \delta_1^-}\\
+ \int_{\frac{3m_R^2}{2k_2^+}}^{+\infty} \frac{d\delta_2^-}{2i\pi} \frac{\left.\text{Disc}_{\delta_2^-}\Sigma_R\left(2k_2^+ \delta_2^- + m_R^2\right)\right|_{\text{1 loop}}}{2k_2^+ \left(\delta_2^-\right)^2 \left(\delta_1^- + \delta_2^- - \Delta_R(\vec{k}_1, \vec{k}_2)\right)}.
\end{multline}
After substituting back into the integral over \(\delta_1^-\), the first term can be transformed again using the dispersion relation. Overall, we find:
\begin{multline}
\left.\psi_{\phi\to\varphi\varphi}(\vec{k}_1, \vec{k}_2)\right|_{\text{Fig.~\ref{fig:two-loop}}+\text{ct}} = \frac{\phi(\vec{k}_1 + \vec{k}_2)}{2(k_1^+ + k_2^+)} \lambda_R \Bigg\{ -\frac{\left(\left.\Sigma_R'(m_R^2)\right|_{\text{1 loop}}\right)^2}{\Delta_R(\vec{k}_1, \vec{k}_2)} \\
+ \left.\Sigma_R'(m_R^2)\right|_{\text{1 loop}} \left(\frac{\left.\Sigma_R\left(2k_1^+\Delta_R(\vec{k}_1, \vec{k}_2) + m_R^2\right)\right|_{\text{1 loop}}}{2k_1^+ \left(\Delta_R(\vec{k}_1, \vec{k}_2)\right)^2} + [\vec{k}_1 \leftrightarrow \vec{k}_2]\right) \\
- \int_{\frac{3m_R^2}{2k_1^+}}^{+\infty} \frac{d\delta_1^-}{2i\pi} \int_{\frac{3m_R^2}{2k_2^+}}^{+\infty} \frac{d\delta_2^-}{2i\pi} \frac{\left.\text{Disc}_{\delta_1^-}\Sigma_R\left(2k_1^+ \delta_1^- + m_R^2\right)\right|_{\text{1 loop}} \left.\text{Disc}_{\delta_2^-}\Sigma_R\left(2k_2^+ \delta_2^- + m_R^2\right)\right|_{\text{1 loop}}}{2k_1^+ \left(\delta_1^-\right)^2 2k_2^+ \left(\delta_2^-\right)^2 \left(\delta_1^- + \delta_2^- - \Delta_R(\vec{k}_1, \vec{k}_2)\right)} \Bigg\}.
\label{eq:relation-scalar-2-loop-symmetric}
\end{multline}

In the formal limit \(\Delta_R \to 0\), the last term in Eq.~\eqref{eq:relation-scalar-2-loop-symmetric} remains finite. The singular terms sum up to
\be
\left.\psi_{\phi\to\varphi\varphi}(\vec{k}_1, \vec{k}_2)\right|_{\text{Fig.~\ref{fig:two-loop}}+\text{ct}}  \underset{\Delta_R \to 0}{\longrightarrow} \frac{\phi(\vec{k}_1 + \vec{k}_2)}{2(k_1^+ + k_2^+)} \lambda_R \frac{\left(\left.\Sigma_R'(m_R^2)\right|_{\text{1 loop}}\right)^2}{\Delta_R(\vec{k}_1, \vec{k}_2)}.
\ee
Adding all self-energy contributions up to order $\bar\lambda_R^4$, the singular parts would still be collectively expressed in the form of Eq.~\eqref{eq:self-energies-onshell-limit}.


\section{Conclusion and outlook}
\label{sec:conclusion}

In this study, we established a framework to derive light-cone wave functions from covariant off-shell amplitudes in scalar field theory. Our approach hinges on a conjectured formula, Eq.~\eqref{eq:relation-scalar} or~\eqref{eq:relation-scalar-integrated-integrated}, that bridges covariant amplitudes and light-cone wave functions, validated through detailed comparisons with one-loop LCPT calculations (Sec.~\ref{sec:validation}). We streamlined the computation of two-particle wave functions of a single asymptotic particle at one-loop accuracy by converting known covariant amplitudes, effectively bypassing the complexities inherent in light-cone quantization methods (Sec.~\ref{sec:re-computation}). We suggested that this framework may be applied to higher orders (Sec.~\ref{sec:two-loops}).

This work sets the stage for future research. First of all, we will extend these methods to gauge theories in an upcoming publication: checking whether our formula, extended to quantum electrodynamics/chromodynamics, is able to reproduce the known $\gamma^*\to q\bar q$ wave-function at one-loop accuracy (see e.g.~\cite{Beuf:2016wdz,Beuf:2017bpd}) would be a good test. Second, it is crucial to explore whether Eq.~\eqref{eq:relation-scalar-integrated-integrated} generalizes beyond the regime in which it has been validated here. Indeed, at this stage, our confidence in Eq.~\eqref{eq:relation-scalar} and~\eqref{eq:relation-scalar-integrated-integrated} rests on its term-by-term agreement with LCPT in a regime where the latter is itself unambiguous and free of the complications alluded to in the Introduction (see also, e.g., Ref.~\cite{Polyzou:2023vjj}). Extending the formula to a regime where LCPT itself is genuinely problematic -- one of the original motivations for this program -- has not yet been demonstrated. Regardless, applying this method to compute higher-order perturbative wave functions, leveraging the sophisticated literature on covariant scattering amplitudes and Feynman diagrams, presents an exciting direction for further study.

Fundamental questions about light-cone quantization, such as the systematic implementation of renormalization, or, even more fundamentally, the equivalence of light-cone and covariant quantizations, have been repeatedly raised. We hope our approach may help shedding new light on some of these issues.

\section*{Acknowledgments}

The author gratefully acknowledges A.-K. Angelopoulou for her curiosity and engagement during her Master’s internship, which marked the beginning of this research project~\cite{Angelopoulou:2021}.


\appendix
\section{One-loop triangle diagram: massless limit results}
\label{sec:appendix-triangle}

For the detailed calculation in the main text, we need the expression for the triangle diagram $\Gamma_{3R}$ in the massless theory when the first argument is zero. Equation~(\ref{eq:Gamma3R}) simplifies as follows:
\begin{equation}
\left.\Gamma_{3R}(0, k_1^2, k_2^2)\right|_{\substack{{{\cal O}(\bar\lambda_R^3)}\\m_R\to 0}} = \frac{\bar\lambda_R^3}{(4\pi)^3} \int_0^1 du \int_0^{1 - u} dv \ln\left[\frac{-(1 - u - v)(uk_1^2 + vk_2^2) - i0^+}{\mu^2}\right].
\end{equation}
Throughout, we assume \(k_1^2\) and \(k_2^2\) are real and not simultaneously zero.

\paragraph{Imaginary part}

\begin{figure}
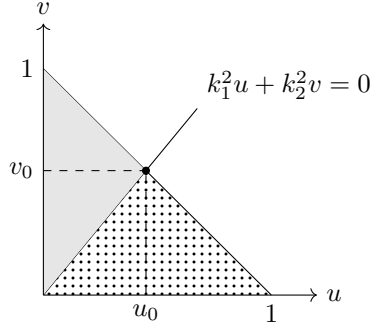

\centerline\uvplane
\caption{Integration variables \((u,v)\) for \(\Gamma_{3R}\). The triangular region defined by the system of equations \(\{0 \leq u \leq 1, 0 \leq v \leq 1 - u\}\) represents the full integration domain . The dotted (respectively gray) subregion indicates the domain contributing to the discontinuity of \(\Gamma_{3R}\) for \(k_1^2 > 0\) (\(k_1^2 < 0\)) and \(k_2^2 < 0\) (\(k_2^2 > 0\)).}
\label{fig:uv-plane}
\end{figure}

The function \(\Gamma_{3R}\) may develop a nonzero imaginary part, defined as
\begin{equation}
\Im\Gamma_{3R} = \frac{\Gamma_{3R} - (\Gamma_{3R})^*}{2i},
\end{equation}
whenever the logarithm in the integrand acquires an imaginary part. For real positive $x$, choosing the principal branch of the complex logarithm, $\ln x$ is real, but
\be
\ln (-x-i0^+)=\lim_{\epsilon\to0^+}\ln\left(xe^{-i(\pi-\epsilon)}\right)=\ln x-i\pi.
\ee
Hence $\Im\Gamma_{3R}\ne 0$ if there exist parameters \(u\) and \(v\) within the integration region such that \(k_1^2u + k_2^2v > 0\). The imaginary part of \(\Gamma_{3R}\) then reads
\be
\Im \left.\Gamma_{3R}(0, k_1^2, k_2^2)\right|_{\substack{{{\cal O}(\bar\lambda_R^3)}\\m_R\to 0}} = \frac{\bar\lambda_R^3}{(4\pi)^3} \int_0^1 du \int_0^{1 - u} dv \,{\mathbbm 1}_{\{k_1^2u + k_2^2v > 0\}}(-\pi).
\ee
Up to the multiplicative factor \((-\pi)\) from the imaginary part of the logarithm, the double integral corresponds to the area in the \((u,v)\)-plane where the integration domain intersects the half-plane defined by \(k_1^2u + k_2^2v \geq 0\). Let us denote by $(u_0,v_0)$ (see Fig.~\ref{fig:uv-plane}) the solution to the linear system
\be
\left\{
\begin{aligned}
&u_0+v_0=1\\
&k_1^2 u_0+k_2^2 v_0=0
\end{aligned}
\right.
\quad\implies\quad
\left(
u_0=\frac{k_2^2}{k_2^2 - k_1^2}, v_0=\frac{k_1^2}{k_1^2 - k_2^2}
\right).
\ee
This area is nonzero in the following cases:
\begin{itemize}
\item If \(k_1^2 > 0\) and \(k_2^2 \leq 0\): the relevant region is the interior of the shaded triangle in Fig.~\ref{fig:uv-plane}, with area $v_0/2$.
\item If \(k_1^2 \leq 0\) and \(k_2^2 > 0\): the relevant region is the dotted triangle in Fig.~\ref{fig:uv-plane}, with area $u_0/2$.
\item If \(k_1^2 \geq 0\), \(k_2^2 \geq 0\), and \(k_1^2\times k_2^2 > 0\): the entire integration domain contributes, yielding an area of \(\frac{1}{2}\).
\end{itemize}
Combining these results, we obtain
\begin{equation}
\Im \left.\Gamma_{3R}(0, k_1^2, k_2^2)\right|_{\substack{{{\cal O}(\bar\lambda_R^3)}\\m_R\to 0}} = -\frac{\bar\lambda_R^3}{(4\pi)^3}\frac{\pi}{2}\left(\frac{k_1^2}{k_1^2 - k_2^2}\mathbbm{1}_{\{k_1^2 > 0, k_2^2 \leq 0\}} + \frac{k_2^2}{k_2^2 - k_1^2}\mathbbm{1}_{\{k_1^2 \leq 0, k_2^2 > 0\}} + \mathbbm{1}_{\{k_1^2 \geq 0, k_2^2 \geq 0\}}\right).
\label{eq:Gamma3R-Im}
\end{equation}
The first two terms correspond to
\begin{equation}
\frac{1}{2i}\text{Disc}_{k_1^2}\left.\Gamma_{3R}(0, k_1^2, k_2^2)\right|_{\substack{\mathcal{O}(\bar\lambda_R^3) \\ m_R \to 0}}
\quad \text{and} \quad
\frac{1}{2i}\text{Disc}_{k_2^2}\left.\Gamma_{3R}(0, k_1^2, k_2^2)\right|_{\substack{\mathcal{O}(\bar\lambda_R^3) \\ m_R \to 0}},
\end{equation}
respectively. These expressions are consistent with the literature: e.g., after adjusting notations and conventions, they agree with Eq.~(B.3) of Ref.~\cite{Abreu:2014cla}.

In principle, this method for computing the imaginary part of \(\Gamma_{3R}\) can be extended to the massive case (\(m_R > 0\)). However, the argument of the logarithm then becomes a quadratic form in the Feynman parameters \((u,v)\), and the area of the relevant region of the plane, namely the intersection of the triangular integration domain with a region bounded by a conic section, is more challenging to evaluate for generic invariants \(k_1^2\), \(k_2^2\).

We note that this approach to computing discontinuities is a simple special case of the general method recently proposed in Ref.~\cite{Britto:2023rig}. The standard approach, by contrast, involves calculating cut Feynman diagrams using Cutkosky rules~\cite{Cutkosky:1960sp}; see e.g. Ref.~\cite{Muhlbauer:2022ylo}.

\paragraph{Expression in the region where the imaginary part vanishes}

When both \(k_1^2\) and \(k_2^2\) are negative, \(\Gamma_{3R}\) is purely real. The integrations over $v$ and $u$ are somewhat tedious, but straightforward. We obtain the following expression:
\begin{equation}
\left.\Gamma_{3R}(0, k_1^2, k_2^2)\right|_{\substack{{{\cal O}(\bar\lambda_R^3)}\\m_R\to 0}} = \frac{\bar\lambda_R^3}{2(4\pi)^3}\left(\frac{k_1^2 \ln(-k_1^2/\mu^2) - k_2^2 \ln(-k_2^2/\mu^2)}{k_1^2 - k_2^2} - 3\right).
\label{eq:Gamma3R-real}
\end{equation}



\end{document}